\pdfoutput=1
   \documentclass[fleqn,usenatbib,usedcolumn]{mnras}
   \usepackage[british]{babel}             
   \let\la=\lesssim 
   \usepackage[T1]{fontenc}                
   \usepackage{graphicx}                   
\usepackage{booktabs}  
\usepackage{pdflscape}   
\usepackage{siunitx}
\usepackage{pgfplots}
\pgfplotsset{compat=newest}

\DeclareSIUnit\mols{{molec.}\,s^{-1}}

   \volume{\textrm{submitted}}
\usepackage{amsmath} 
\usepackage{amssymb} 

\def\rh{$r_{\mathrm{H}}$}

\title[Gemini and Lowell Observations of 67P]{Gemini and Lowell Observations of 67P/Churyumov-Gerasimenko During the Rosetta Mission}

\author[M.M. Knight et al.]{%
Matthew M. Knight,$^{1,2}$\thanks{E-mail: mmk8a@astro.umd.edu (MMK)}
Colin Snodgrass,$^{3}$
Jean-Baptiste Vincent,$^{4}$
Blair C.~Conn,$^{5,6}$
\newauthor
Brian A. Skiff,$^{2}$
David G.~Schleicher,$^{2}$
and Tim Lister$^{7}$
\\
\smallskip\\
$^{1}$Department of Astronomy, University of Maryland, College Park, MD 20742, USA\\
$^{2}$Lowell Observatory, 1400 W. Mars Hill Rd, Flagstaff, AZ 86001, USA\\
$^{3}$School of Physical Sciences, The Open University, Walton Hall, Milton Keynes, MK7 6AA, UK\\
$^{4}$DLR Institute of Planetary Research, Berlin, Germany\\
$^{5}$Research School of Astronomy and Astrophysics, Australian National University, Canberra, ACT 2611, Australia\\
$^{6}$Gemini Observatory, Casilla 603, La Serena, Chile\\
$^{7}$Las Cumbres Observatory Global Telescope Network, 6740 Cortona Drive Suite. 102, Goleta, CA 93117, USA
}

\date{Accepted 2017 September 19. Received 2017 September 18; in original form 2017 April 20}

\pubyear{2017}

\begin{document}
\label{firstpage}
\pagerange{\pageref{firstpage}--\pageref{lastpage}}
\maketitle

\begin{abstract}
We present observations of comet 67P/Churyumov-Gerasimenko acquired in support of the {\it Rosetta} mission. We obtained usable data on 68 nights from 2014 September until 2016 May, with data acquired regularly whenever the comet was observable. We collected an extensive set of near-IR $J$, $H$, and $Ks$ data throughout the apparition plus visible-light images in $g'$, $r'$, $i'$, and $z'$ when the comet was fainter. We also obtained broadband $R$ and narrowband {\it CN} filter observations when the comet was brightest using telescopes at Lowell Observatory. The appearance was dominated by a central condensation and the tail until 2015 June. From 2015 August onwards there were clear asymmetries in the coma, which enhancements revealed to be due to the presence of up to three features (i.e., jets). The features were similar in all broadband filters; {\it CN} images did not show these features but were instead broadly enhanced in the southeastern hemisphere. Modeling using the parameters from \citet{vincent13} replicated the dust morphology reasonably well, indicating that the pole orientation and locations of active areas have been relatively unchanged over at least the last three apparitions. The dust production, as measured by $A(0^{\circ})f{\rho}$ peaked $\sim$30 days after perihelion and was consistent with predictions from previous apparitions. $A(0^{\circ})f{\rho}$ as a function of heliocentric distance was well fit by a power-law with slope $-$4.2 from 35--120 days post-perihelion. We detected photometric evidence of apparent outbursts on 2015 August 22 and 2015 September 19, although neither was discernible morphologically in this dataset. 
\end{abstract}




\section{Introduction}
ESA's {\it Rosetta} mission to comet 67P/Churyumov-Gerasimenko (henceforth 67P) officially reached the comet on 2014 August 6 at a heliocentric distance (\rh{}) of 3.60~AU prior to perihelion \citep{taylor15}. It followed the comet through perihelion (2015 August 13 at \rh{}~=~1.24~AU) and continued orbiting until the spacecraft was intentionally landed on the nucleus' surface on 2016 September 30 at \rh{}~=~3.83~AU. This 2+ year rendezvous created the first opportunity to study a comet simultaneously in situ and from the Earth over an extended time period. The mission's value is amplified by fully incorporating the Earth-based observations because the lessons learned from 67P during {\it Rosetta} can be extended to far more comets than will be visited by spacecraft in the foreseeable future.

Due to {\it Rosetta}'s location within 67P's coma, it probed vastly different scales (generally $<10^2$~km) than Earth-based observations ($10^{3}-10^{5}$~km), and a full picture of the comet's behavior during the apparition could be gained only by supplementing the spacecraft data with Earth-based observations. Throughout most of the appartion, 67P was difficult to observe from Earth, either because it was distant and faint or because it was near solar conjunction and therefore poorly placed for observing. Its observability from any one location was also limited because 67P was primarily a southern hemisphere object prior to perihelion and a northern hemisphere object post-perihelion.

We undertook a campaign to monitor 67P's dust coma evolution throughout the {\it Rosetta} encounter at visible-light and near-IR wavelengths. Such a long duration campaign necessitated a large telescope for most of the observations. Gemini Observatory proved ideally suited for these observations: its has twin 8-m southern and northern facilities that are equipped with similar  instrumentation, allowing us to acquire a nearly homogeneous dataset throughout the encounter. Furthermore, Gemini's queued scheduling permitted regular observations, even when 67P was at low solar elongations and available only during twilight. Near perihelion, when 67P was brightest and accessible to smaller telescopes, we acquired visible-light images at Lowell Observatory rather than with Gemini.

In an effort to obtain data on an approximately regular cadence, observations were acquired on many nights during poor conditions, either because of weather and atmospheric conditions, or because the comet was observable only during twilight at high airmass. We also observed more intensively around key points in the {\it Rosetta} mission timeline such as the {\it Philae} landing (2014 November 12) and perihelion, both of which occurred during poor observing circumstances from Earth. Thus, our observations focussed primarily on imaging to emphasize monitoring the dust coma morphology, which does not require cloud-free photometric conditions. Most observations were short, snapshot images of the comet; longer duration near-IR spectroscopy was also obtained on three nights when 67P was brightest but they are beyond the scope of the current analysis. In total, we collected usable data on 68 nights from September 2014 through May 2016. 

Preliminary results from this observing campaign have already been published in synopses of 67P's distant activity as observed with large telescopes \citep{snodgrass16} and of the worldwide ground-based observation campaign \citep{snodgrass17}. Here we present comprehensive analysis of our observations. In Section~\ref{sec:obs} we describe the observations and data reductions. We present our results and analysis in Section~\ref{sec:analysis}, focussing on dust coma morphology since our data set is capable of demonstrating 67P's long-term morphological behavior throughout the {\it Rosetta} mission, but also considering the photometric evolution. Finally, in Section~\ref{sec:disc} we discuss how these observations compare to Earth-based observations from previous apparitions and consider connections between remote sensing and in situ studies conducted by {\it Rosetta}.

\section{Observations and Reductions} 
\label{sec:obs}
The imaging observations are summarized in Table~\ref{t:imaging_circ}. Each of the various telescope and instrument combinations is discussed in  its own subsection, below. Owing to 67P's challenging viewing geometry during much of the apparition, observations were frequently acquired at airmasses $>$1.5, during twilight, and at less than optimal atmospheric conditions in order to ensure frequent monitoring of the comet. Nonetheless, typical seeing was $<$1.0 arcsec for all Gemini and Discovery Channel Telescope observations. Seeing at the Lowell 31-in site is $\sim$1~arcsec, but the effective seeing due to dome effects, tracking wobbles, etc.\ was typically 3--5~arcsec.

\subsection{FLAMINGOS-2}
FLAMINGOS-2 \citep{eikenberry04} is a near-IR imager and multi-object spectrograph on Gemini-South. We used it during the 2014B and 2015A semesters in imaging mode where it has a 6.1~arcmin diameter circular field of view and a Hawaii-II (HgCdTe) chip with 2048$\times$2048 pixels, resulting in 0.18~arcsec pixels. We used $J$, $H$, and $Ks$ filters and guided on the comet's ephemeris without using adaptive optics (AO). All observations were acquired in queue mode. We followed standard near-IR observing practices to build deeper integrations by collecting multiple short observations with dithers between them, but did not coadd on chip before reading out. We used the bright read mode to reduce overhead since the extra read noise was dwarfed by the high sky background. The number of exposures and exposure times varied with the comet's brightness as well as with the sky background, but a minimum of five exposures were acquired with a particular filter on a given night. We planned to acquire all three filters on all nights, but occasionally the set was truncated by the telescope operator due to a change in observing conditions, rising sky background, or technical problems. Calibration data (dome flats, darks, short darks) were acquired as part of Gemini's routine observations. The data were reduced in IRAF using the {\it gemini/f2} package and following reduction scripts provided by Gemini\footnote{\href{http://www.gemini.edu/sciops/data/IRAFdoc/f2_imaging_example.cl}{http://www.gemini.edu/sciops/data/IRAFdoc/{\newline}f2\_imaging\_example.cl}}. Sky frames were created from the dithered comet frames since the comet did not fill a significant portion of the field of view. Although the dithers were larger than than the apparent extent of the coma, some low-level coma was likely still present in the sky image, resulting in slight over-removal of the background. This would have impacted photometric measurements (none are presented here), but had minimal effect on our morphological studies as we confirmed that there was no evidence of residual structure in the sky frames. The sky frames were subtracted from the individual images and all images in a given filter on a night were then coadded by centroiding on the comet to produce a single master image.

\subsection{NIRI}
When 67P became a northern hemisphere target (semesters 2015B and 2016A) we used the Near InfraRed Imager and spectrograph \citep[NIRI;][]{hodapp03} on Gemini-North in imaging mode and with queued scheduling. We used NIRI at f/6 where its 1024$\times$1024 pixels cover a square field of view 120$\times$120~arcsec on a side, resulting in 0.117~arcsec pixels. Our observation strategy mimicked the FLAMINGOS-2 observations just described with notable differences being that we used ``medium background'' read mode which is recommended for $J$, $H$, $Ks$ broadband imaging and, when possible, used the same exposure time in all filters to minimize

\begin{landscape}
\begin{table}
	\centering
	\setlength{\tabcolsep}{0.04in}
	\caption{Summary of 67P/Churyumov-Gerasimenko observations and geometric parameters for imaging.$^{a}$}
	\label{t:imaging_circ}
	\begin{tabular}{lccccccccccccccccccl}
		\hline
\multicolumn{1}{c}{Date}&Time&Tel.$^{b}$&Instrument&\multicolumn{9}{c}{Image details (\# of exposures $\times$ Exposure time [s])}&$\Delta$T$^{c}$&$r_\mathrm{H}$$^{d}$&$\mathit{\Delta}$$^{e}$&$\theta$$^{f}$&P.A.$^{g}$&Conditions$^{h}$\\
\cmidrule(lr){5-13}
\multicolumn{1}{c}{(UT)}&(UT)&&&$g'$&$r'$&$i'$&$z'$&$R$&{\it CN}&$J$&$H$&$Ks$&(days)&(AU)&(AU)&($^\circ$)&($^\circ$)&\\		
		\hline
2014 Sep 19--20&23:56--00:58&GS&Flamingos-2&-&-&-&-&-&-&15$\times$45&44$\times$10&17$\times$20&$-$327.2&3.335&2.989&17.2&94&IQ70, CC50\\
2014 Sep 20--21&23:33--00:02&GS&GMOS&1$\times$60&15$\times$60&1$\times$60&1$\times$60&-&-&-&-&-&$-$326.2&3.329&2.997&17.3&94&IQ70, CC50\\
2014 Oct 29&00:16--01:12&GS&Flamingos-2&-&-&-&-&-&-&5$\times$30&44$\times$6&17$\times$15&$-$287.2&3.087&3.315&17.4&89&IQ70, CC50\\
2014 Oct 29&01:17--01:48&GS&GMOS&1$\times$60&15$\times$60&1$\times$60&1$\times$60&-&-&-&-&-&$-$287.2&3.087&3.316&17.4&89&IQ70, CC50\\
2014 Nov 11&00:01--00:24&GS&GMOS&-&15$\times$60&-&-&-&-&-&-&-&$-$275.2&3.001&3.400&16.3&87&IQ70, CC50\\
2014 Nov 12.&00:03--00:27&GS&GMOS&-&15$\times$60&-&-&-&-&-&-&-&$-$274.2&2.995&3.406&16.2&86&IQ85, CC50\\
2014 Nov 13&00:55--01:19&GS&GMOS&-&15$\times$60&-&-&-&-&-&-&-&$-$273.2&2.988&3.412&16.1&86&IQ70, CC50\\
2014 Nov 14&00:13--01:16&GS&Flamingos-2&-&-&-&-&-&-&5$\times$60&44$\times$12&17$\times$5&$-$272.2&2.981&3.418&16.0&86&IQ70, CC50\\
2014 Nov 15&00:29--00:34&GS&GMOS&1$\times$60&1$\times$60&1$\times$60&1$\times$60&-&-&-&-&-&$-$271.2&2.974&3.423&15.9&86&IQ70, CC50\\
2014 Nov 16&00:14--00:38&GS&GMOS&-&15$\times$60&-&-&-&-&-&-&-&$-$270.2&2.968&3.428&15.8&86&IQ70, CC50\\
2014 Nov 17&00:23--00:47&GS&GMOS&-&15$\times$60&-&-&-&-&-&-&-&$-$269.2&2.961&3.433&15.7&85&IQ70, CC50\\
2014 Nov 18&00:05--00:58&GS&Flamingos-2&-&-&-&-&-&-&10$\times$30&44$\times$12&16$\times$15&$-$268.2&2.954&3.438&15.5&85&IQ85, CC50\\
2014 Nov 18&01:02--01:26&GS&GMOS&-&15$\times$60&-&-&-&-&-&-&-&$-$268.2&2.954&3.439&15.5&85&IQ85, CC50\\
2014 Nov 19&00:23--01:08&GS&GMOS&1$\times$60&16$\times$60&1$\times$60&1$\times$60&-&-&-&-&-&$-$267.2&2.947&3.443&15.4&85&IQ70, CC50\\
2015 Jun 20&10:26--10:38&GS&Flamingos-2&-&-&-&-&-&-&5$\times$15&15$\times$13&-&$-$53.8&1.404&1.989&28.8&66&IQ70, CC70\\
2015 Jun 24&10:12--10:38&GS&Flamingos-2&-&-&-&-&-&-&5$\times$15&15$\times$13&-&$-$49.8&1.382&1.960&29.4&67&IQ70, CC70\\
2015 Jun 26&10:26--10:50&GS&Flamingos-2&-&-&-&-&-&-&5$\times$15&12$\times$6&21$\times$10&$-$47.8&1.372&1.946&29.8&68&IQ70, CC70\\
2015 Jun 30&10:22--10:44&GS&Flamingos-2&-&-&-&-&-&-&5$\times$15&15$\times$6&21$\times$10&$-$43.8&1.353&1.920&30.4&69&IQ70, CC50\\
2015 Aug 4&14:56--15:16&GN&NIRI&-&-&-&-&-&-&5$\times$60&15$\times$20&8$\times$45&$-$8.6&1.248&1.782&33.6&82&IQ20, CC50\\
2015 Aug 8&15:18--15:36&GN&NIRI&-&-&-&-&-&-&5$\times$60&15$\times$20&4$\times$22.5&$-$4.6&1.244&1.776&33.8&83&IQ70, CC70\\
2015 Aug 12&14:49--15:16&GN&NIRI&-&-&-&-&-&-&5$\times$60&15$\times$20&9$\times$45&$-$0.6&1.243&1.772&33.9&85&IQ70, CC70\\
2015 Aug 18&11:25--12:00&31in&CCD&-&-&-&-&9$\times$60&9$\times$180&-&-&-&$+$5.3&1.245&1.769&33.9&88&Thin clouds\\
2015 Aug 19&11:36--12:33&31in&CCD&-&-&-&-&34$\times$60&3$\times$180&-&-&-&$+$6.3&1.246&1.768&34.0&88&Photometric\\
2015 Aug 20&11:26--11:57&31in&CCD&-&-&-&-&9$\times$60&6$\times$180&-&-&-&$+$7.3&1.247&1.768&34.0&89&Photometric\\
2015 Aug 20&14:49--15:26&GN&NIRI&-&-&-&-&-&-&7$\times$60&12$\times$20&16$\times$25&$+$7.4&1.247&1.768&34.0&89&IQ70, CC50\\
2015 Aug 21&11:21--12:02&31in&CCD&-&-&-&-&27$\times$60&3$\times$180&-&-&-&$+$8.3&1.248&1.768&34.0&89&Photometric\\
2015 Aug 22&11:17--11:56&31in&CCD&-&-&-&-&9$\times$60&9$\times$180&-&-&-&$+$9.3&1.249&1.768&34.0&90&Thin clouds\\
2015 Aug 23&11:17--11:59&31in&CCD&-&-&-&-&28$\times$60&3$\times$180&-&-&-&$+$10.3&1.250&1.768&34.0&90&Thin clouds\\
2015 Aug 26&15:06--15:28&GN&NIRI&-&-&-&-&-&-&7$\times$60&15$\times$20&14$\times$25&$+$13.4&1.254&1.769&34.0&91&IQ20, CC70\\
2015 Aug 29&11:23--11:58&31in&CCD&-&-&-&-&9$\times$60&9$\times$180&-&-&-&$+$16.3&1.260&1.770&33.9&93&Clouds\\
2015 Sep 12&11:23--12:18&31in&CCD&-&-&-&-&12$\times$30&12$\times$180&-&-&-&$+$30.3&1.298&1.780&33.8&98&Photometric\\
2015 Sep 15&15:13--15:36&GN&NIRI&-&-&-&-&-&-&10$\times$20&15$\times$20&-&$+$33.4&1.310&1.783&33.7&100&IQ20, CC50\\
2015 Sep 16&15:19--15:43&GN&NIRI&-&-&-&-&-&-&10$\times$20&15$\times$20&-&$+$34.4&1.313&1.784&33.7&100&IQ20, CC70\\
2015 Sep 18&11:38--12:17&31in&CCD&-&-&-&-&9$\times$60&9$\times$180&-&-&-&$+$36.3&1.321&1.786&33.7&101&Photometric\\
2015 Sep 19&11:42--12:24&31in&CCD&-&-&-&-&28$\times$60&3$\times$180&-&-&-&$+$37.3&1.325&1.787&33.7&101&Photometric\\
2015 Sep 19&15:03--15:24&GN&NIRI&-&-&-&-&-&-&10$\times$20&15$\times$20&15$\times$20&$+$37.4&1.325&1.787&33.7&101&IQ70, CC50\\
2015 Sep 20&11:41--12:20&31in&CCD&-&-&-&-&9$\times$60&9$\times$180&-&-&-&$+$38.3&1.329&1.788&33.7&101&Photometric\\
2015 Sep 23&11:13--12:16&DCT&LMI&-&-&-&-&6$\times$30&6$\times$180&-&-&-&$+$41.3&1.342&1.791&33.6&102&Photometric\\
2015 Sep 24&11:07--12:11&DCT&LMI&-&-&-&-&6$\times$30&6$\times$180&-&-&-&$+$42.3&1.347&1.792&33.6&103&Photometric\\
2015 Sep 24&11:51--12:30&31in&CCD&-&-&-&-&12$\times$30&12$\times$180&-&-&-&$+$42.3&1.347&1.792&33.6&103&Photometric\\
2015 Sep 25&11:51--12:19&31in&CCD&-&-&-&-&6$\times$30&6$\times$180&-&-&-&$+$43.3&1.352&1.793&33.6&103&Photometric\\
2015 Sep 26&11:53--12:16&31in&CCD&-&-&-&-&6$\times$30&6$\times$180&-&-&-&$+$44.3&1.356&1.794&33.6&103&Photometric\\
2015 Sep 30&11:52--12:32&31in&CCD&-&-&-&-&9$\times$30&9$\times$180&-&-&-&$+$48.3&1.376&1.798&33.6&105&Thin clouds\\
2015 Oct 2&11:54--12:29&31in&CCD&-&-&-&-&9$\times$30&9$\times$180&-&-&-&$+$50.3&1.386&1.800&33.5&105&Thin clouds\\
2015 Oct 3&11:52--12:33&31in&CCD&-&-&-&-&9$\times$30&9$\times$180&-&-&-&$+$51.3&1.392&1.801&33.5&106&Photometric\\

		\hline
	\end{tabular}
\end{table}
\end{landscape}

\begin{landscape}
\begin{table}
	\centering
	\setlength{\tabcolsep}{0.04in}
	\contcaption{Summary of 67P/Churyumov-Gerasimenko observations and geometric parameters for imaging.$^{a}$}
	\label{t:imaging_circ_contd}
	\begin{tabular}{lccccccccccccccccccl}
		\hline
\multicolumn{1}{c}{Date}&Time&Tel.$^{b}$&Instrument&\multicolumn{9}{c}{Image details (\# of exposures $\times$ Exposure time [s])}&$\Delta$T$^{c}$&$r_\mathrm{H}$$^{d}$&$\mathit{\Delta}$$^{e}$&$\theta$$^{f}$&P.A.$^{g}$&Conditions$^{h}$\\
\cmidrule(lr){5-13}
\multicolumn{1}{c}{(UT)}&(UT)&&\phantom{Flamingos-2}&$g'$&$r'$&$i'$&$z'$&$R$&{\it CN}&$J$&$H$&$Ks$&(days)&(AU)&(AU)&($^\circ$)&($^\circ$)&\\		
		\hline
2015 Oct 4\phantom{9--20}&12:06--12:29&31in&CCD&-&-&-&-&6$\times$30&6$\times$180&-&-&-&$+$52.3&1.397&1.802&33.5&106&Thin clouds\\
2015 Oct 10&14:57--15:18&GN&NIRI&-&-&-&-&-&-&10$\times$20&15$\times$20&15$\times$20&$+$58.4&1.431&1.806&33.5&108&IQ20, CC50\\
2015 Oct 13&11:52--12:03&31in&CCD&-&-&-&-&10$\times$60&-&-&-&-&$+$61.3&1.448&1.807&33.4&109&Thin clouds\\
2015 Oct 15&14:49--15:10&GN&NIRI&-&-&-&-&-&-&10$\times$20&15$\times$20&15$\times$20&$+$63.4&1.461&1.808&33.4&109&IQ70, CC50\\
2015 Oct 26&15:11--15:33&GN&NIRI&-&-&-&-&-&-&10$\times$20&15$\times$20&15$\times$20&$+$74.4&1.530&1.808&33.4&112&IQ70, CC70\\
2015 Nov 1&15:41--15:58&GN&NIRI&-&-&-&-&-&-&10$\times$20&15$\times$20&-&$+$80.4&1.571&1.804&33.3&113&IQ70, CC50\\
2015 Nov 2&12:38--12:49&31in&CCD&-&-&-&-&10$\times$60&-&-&-&-&$+$81.3&1.577&1.804&33.3&114&Photometric\\
2015 Nov 2&15:17--15:24&GN&NIRI&-&-&-&-&-&-&-&-&15$\times$20&$+$81.4&1.577&1.804&33.3&114&IQ70, CC50\\
2015 Nov 8&12:40--12:52&31in&CCD&-&-&-&-&10$\times$60&-&-&-&-&$+$87.3&1.619&1.797&33.2&115&Thin clouds\\
2015 Nov 9&12:41--12:52&31in&CCD&-&-&-&-&10$\times$60&-&-&-&-&$+$88.3&1.626&1.796&33.2&115&Photometric\\
2015 Nov 10&12:42--12:53&31in&CCD&-&-&-&-&10$\times$60&-&-&-&-&$+$89.3&1.633&1.794&33.2&115&Photometric\\
2015 Nov 18&12:46--12:57&31in&CCD&-&-&-&-&10$\times$60&-&-&-&-&$+$97.3&1.690&1.780&33.0&117&Thin clouds\\
2015 Nov 19&12:52--13:03&31in&CCD&-&-&-&-&10$\times$60&-&-&-&-&$+$98.3&1.698&1.778&32.9&117&Photometric\\
2015 Nov 20&12:51--13:02&31in&CCD&-&-&-&-&10$\times$60&-&-&-&-&$+$99.3&1.705&1.776&32.9&118&Photometric\\
2015 Nov 27&12:47--12:58&31in&CCD&-&-&-&-&10$\times$60&-&-&-&-&$+$106.3&1.757&1.758&32.6&119&Photometric\\
2015 Nov 27&15:00--15:21&GN&NIRI&-&-&-&-&-&-&10$\times$20&15$\times$20&15$\times$20&$+$106.4&1.758&1.758&32.6&119&IQ85, CC70\\
2015 Nov 30&12:30--12:41&31in&CCD&-&-&-&-&10$\times$60&-&-&-&-&$+$109.3&1.779&1.749&32.5&119&Photometric\\
2015 Dec 1&12:31--12:42&31in&CCD&-&-&-&-&10$\times$60&-&-&-&-&$+$110.3&1.787&1.746&32.4&120&Thin clouds\\
2015 Dec 6&11:59--12:52&DCT&LMI&-&-&-&-&6$\times$180&3$\times$600&-&-&-&$+$115.3&1.825&1.730&32.1&121&Cirrus\\
2015 Dec 6&15:20--15:52&GN&NIRI&-&-&-&-&-&-&15$\times$20&15$\times$20&18$\times$20&$+$115.4&1.826&1.729&32.0&121&IQ20, CC50\\
2015 Dec 7&12:19--12:27&DCT&LMI&-&-&-&-&3$\times$120&-&-&-&-&$+$116.3&1.832&1.726&32.0&121&Photometric\\
2015 Dec 18&15:34--16:12&GN&NIRI&-&-&-&-&-&-&22$\times$20&46$\times$20&-&$+$127.4&1.917&1.683&30.9&123&IQ85, CC70\\
2015 Dec 24&14:46--15:36&GN&NIRI&-&-&-&-&-&-&22$\times$20&45$\times$20&27$\times$20&$+$133.4&1.963&1.658&30.0&123&IQ70, CC50\\
2016 Jan 3&15:07--16:06&GN&NIRI&-&-&-&-&-&-&22$\times$20&45$\times$20&27$\times$20&$+$143.4&2.041&1.613&28.3&124&IQ70, CC70\\
2016 Jan 19&11:41--13:21&GN&NIRI&-&-&-&-&-&-&30$\times$20&73$\times$20&55$\times$20&$+$159.3&2.163&1.544&24.2&125&IQ70, CC50\\
2016 Jan 31&15:01--16:16&GN&NIRI&-&-&-&-&-&-&30$\times$20&73$\times$20&55$\times$20&$+$171.4&2.256&1.504&19.9&125&IQ85, CC50\\
2016 Feb 16&11:00--12:01&GN&NIRI&-&-&-&-&-&-&17$\times$30&57$\times$20&28$\times$30&$+$187.3&2.376&1.485&13.1&123&IQ70, CC50\\
2016 Feb 16&12:05--12:25&GN&GMOS&3$\times$60&3$\times$60&3$\times$60&3$\times$60&-&-&-&-&-&$+$187.3&2.376&1.485&13.0&123&IQ70, CC50\\
2016 Mar 10&10:09--11:09&GN&NIRI&-&-&-&-&-&-&17$\times$30&57$\times$20&28$\times$30&$+$210.2&2.547&1.563&3.9&120&IQ85, CC50\\
2016 Mar 10&11:22--11:48&GN&GMOS&5$\times$60&-&-&-&-&-&-&-&-&$+$210.2&2.547&1.563&3.9&120&IQ85, CC50\\
2016 Apr 13&06:09--06:56&GN&GMOS&3$\times$60&18$\times$60&3$\times$60&3$\times$60&-&-&-&-&-&$+$244.1&2.790&1.941&13.1&117&IQ85, CC50\\
2016 Apr 30&05:53--06:12&GN&GMOS&3$\times$60&3$\times$60&3$\times$60&3$\times$60&-&-&-&-&-&$+$261.0&2.907&2.225&16.8&116&IQ85, CC50\\
2016 Apr 30&06:21--07:20&GN&NIRI&-&-&-&-&-&-&17$\times$30&57$\times$20&28$\times$30&$+$261.0&2.907&2.225&16.7&116&IQ85, CC50\\
2016 May 23&06:12--06:30&GN&GMOS&3$\times$60&3$\times$60&3$\times$60&3$\times$60&-&-&-&-&-&$+$284.0&3.062&2.667&18.8&116&IQ70, CC50\\
2016 May 23&06:36--07:35&GN&NIRI&-&-&-&-&-&-&17$\times$30&57$\times$20&28$\times$30&$+$284.0&3.062&2.666&18.8&116&IQ70, CC50\\
2016 May 28&06:44--07:16&GN&GMOS&-&15$\times$60&-&-&-&-&-&-&-&$+$289.1&3.095&2.768&18.9&116&IQ70, CC70\\

		\hline
\multicolumn{19}{l}{$^{a}$ Geometric parameters taken at the midpoint of each night's observations.}\\		
\multicolumn{19}{l}{$^{b}$ Telescope: GS = Gemini South, GN = Gemini North, DCT = Discovery Channel Telescope, 31in = Lowell 31-in.}\\
\multicolumn{19}{l}{$^{c}$ Time relative to perihelion.}\\
\multicolumn{19}{l}{$^{d}$ Heliocentric distance.}\\
\multicolumn{19}{l}{$^{e}$ Earth-comet distance.}\\
\multicolumn{19}{l}{$^{f}$ Solar phase angle.}\\
\multicolumn{19}{l}{$^{g}$ Position angle of the Sun, measured from north through east.}\\
\multicolumn{19}{l}{$^{h}$ Gemini codes generated automatically at the time of observation by the seeing monitor: IQ = image quality, CC = cloud cover. Number represents percentage}\\ 
\multicolumn{19}{l}{$^{\phantom{h}}$ of time across all nights that conditions are at least that good (lower numbers are better).}\\
	\end{tabular}
\end{table}
\end{landscape}

\noindent  the ``first frame issue'' that occurs whenever the detector configuration changes (read mode, exposure time, etc.) and causes poor background subtraction in the subsequent image. The data were reduced in IRAF using a combination of the {\it gemini/niri} package following reduction scripts provided by Gemini\footnote{\href{http://www.gemini.edu/sciops/data/IRAFdoc/niri_imaging_example.cl}{http://www.gemini.edu/sciops/data/IRAFdoc/{\newline}niri\_imaging\_example.cl}} and additional IRAF and Python scripts provided by A.~Stephens (Gemini).

\subsection{GMOS-South and -North}
We observed with both Gemini Multi-Object Spectrographs \citep[GMOS;][]{hook04} in imaging mode, using GMOS-South (GMOS-S) in semesters 2014B and 2015A and GMOS-North (GMOS-N) in semester 2016A. Each GMOS has a square field of view 5.5~arcmin on a side. GMOS-S uses the new, red-sensitive Hamamatsu detector \citep{gimeno16} with 6266$\times$4176 pixels and an image scale of 0.08 arcsec pixel$^{-1}$, while GMOS-N uses the e2v DD chip having 6144$\times$4608 pixels and an image scale of 0.0728 arcsec pixel$^{-1}$. All observations were binned 2$\times$2 resulting in 0.160 arcsec (GMOS-S) and 0.1456 arcsec (GMOS-N) pixels. Each detector has three chips with gaps between them, but as the position of 67P was well known and it was not highly extended when we observed it, the comet was always located on the middle chip and did not extend significantly into the chip gaps. Images were guided on the comet's ephemeris without using AO. 

We had two types of observation with GMOS: short $g'$, $r'$, $i'$, and $z'$ color sequences and deeper $r'$ sequences for morphology. All images used the slow read mode, and were dithered whenever multiple images were acquired. The number of exposures varied with the comet's brightness; 1--3 images were acquired per filter for color sequences, while the deeper $r'$ sequences consisted of 5--18 images. The data were reduced using the {\it gemini/gmos} package and following the reduction scripts provided by Gemini\footnote{\href{http://www.gemini.edu/sciops/data/IRAFdoc/gmos_imaging_example.cl}{http://www.gemini.edu/sciops/data/IRAFdoc/{\newline}gmos\_imaging\_example.cl}} to remove bias and perform flat fielding. Due to the snapshot nature of the color sequences, there were insufficient $z'$ images to remove fringing. However, fringing had a negligible effect on our photometry because it is small for GMOS-S ($\sim$1\% of the background\footnote{\href{http://www.gemini.edu/sciops/instruments/gmos/imaging/fringing/gmossouth}{http://www.gemini.edu/sciops/instruments/gmos/imaging/{\newline}fringing/gmossouth}}) when the comet was fainter, while the somewhat larger GMOS-N fringing ($\sim$2.5\% of the background\footnote{\href{https://www.gemini.edu/node/10648}{https://www.gemini.edu/node/10648}}) was offset by the comet's much brighter appearance. The data were photometrically calibrated using the Gemini zero points and extinction coefficients; for GMOS-S calibrations where a color term is also needed, we assumed solar color. For GMOS data taken at low galactic latitude in 2014, we also employed Difference Image Analysis techniques, implemented in IDL, to remove background stars \citep{bramich08}.

\subsection{Discovery Channel Telescope (DCT)}
We utilized the Large Monolithic Imager \citep[LMI;][]{massey13} on Lowell Observatory's 4.3-m Discovery Channel Telescope (DCT) on four nights. LMI has an e2v CCD with a square field of view 12.3 arcmin on a side. It contains 6256$\times$6160 pixels which we binned on chip 2$\times$2, yielding 0.24 arcsec pixels. We used a broadband Cousins {\it R}-band filter and a comet HB narrowband {\it CN} filter (\citealt{farnham00}; ``HB'' is the formal name of the filter set). The comet was observed near an airmass of 1.9 and during astronomical twilight in 2015 September, but near an airmass of 1.3 and during dark time in 2015 December. The telescope was guided on the comet's ephemeris. Images were bias- and flat-field-corrected in IDL following standard procedures \citep[e.g.,][]{knight15a}. $R$-band images were absolutely calibrated using field stars in the UCAC4 catalog \citep{zacharias13}. Results were close to the absolute calibrations performed using \citet{landolt09} standard stars on two of the nights. {\it CN} absolute calibrations were not performed as the appropriate standard stars were not observed since the comet observations were acquired primarily to assess gas coma morphology.

\subsection{31-in}
We obtained useful data on 36 nights using Lowell Observatory's 31-in (0.8-m) telescope in robotic mode. Although we refer to it as the ``31-in'' throughout this paper because that is its formal name, we note that the primary is stopped down to 29-in and the telescope is therefore effectively a 0.7-m. The 31-in has an e2v CCD42-40 chip with 2138$\times$2052 pixels and a square field of view 15.7~arcmin on a side, yielding 0.456 arcsec pixels. We used a broadband Cousins {\it R}-band filter and the HB narrowband {\it CN} filter, and tracked on the comet's ephemeris. As shown in Table~\ref{t:imaging_circ}, conditions varied from night to night but were often non-photometric. Effective seeing was typically 3--5~arcsec, and many images were acquired during astronomical twilight. Images were bias- and flat-field-corrected in IDL following standard procedures \citep[e.g.,][]{knight15a}. We photometrically calibrated the data each night using UCAC4 catalog field stars \citep{zacharias13} and confirmed that the results were close to the typical calibration coefficients used for this telescope, demonstrating that cirrus was minimal (usually $<$0.1~mag, always $<$0.35~mag) on the nights used for this analysis.

 \section{RESULTS AND ANALYSIS} 
\label{sec:analysis}
\subsection{Morphology Assessment}
\subsubsection{Dust Morphology}
\begin{figure*}
  \centering
 \includegraphics[width=175mm]{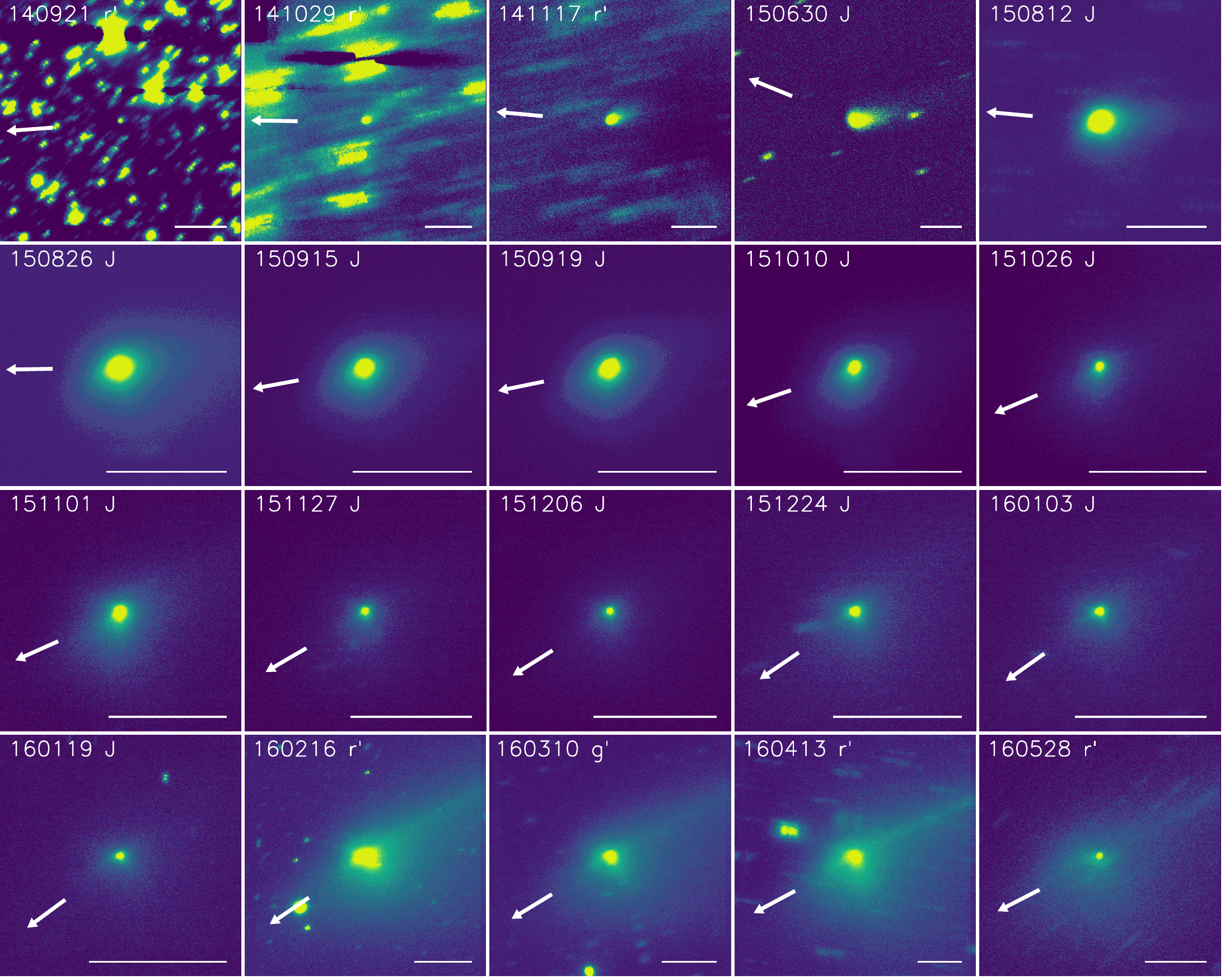}
  \caption[Composite of unenhanced images]{Evolution of 67P's morphology from 2014 September through 2016 May. The date (YYMMDD format), filter, and a scale bar 30,000~km across are given on each panel. All images are centered on the comet. North is up and east is to the left. The direction to the Sun is indicated by the arrow on each panel. The color scale varies from image to image, but yellow is bright and blue/black is faint. Trailed stars are visible in some frames.}
  \label{fig:dust_unenhanced}
\end{figure*}

As previously discussed, our primary goal was assessment of coma morphology. To improve signal-to-noise in the coma, all images in the same filter on a given night were centroided and combined to create a single deeper exposure for the night. The vast majority of our images were acquired in bandpasses typically dominated by dust: $R$, $r'$, $i'$, $z'$, $J$, $H$, and $Ks$. Spectroscopy (e.g., Figure 1 of \citealt{feldman04}) reveals that in most comets the $g'$ filter has some gas contamination (C$_3$ and C$_2$), however, both appear to be relatively low in 67P (e.g., \citealt{schleicher06b,opitom17}). Furthermore, as shown below, the morphology in this filter suggests gas contamination is minimal. Unsurprisingly, the bulk morphology looked similar in all of these filters at a given epoch. As shown in Fig.~\ref{fig:dust_unenhanced}, the general morphology evolved over the apparition. 67P's initial appearance in late 2014 was faint with a central condensation and a weak tail, and it changed little through our last observations before solar conjunction in 2015 June. From 2015 August through 2016 January, when 67P was closest to Earth and brightest, the extended coma was obvious and asymmetries in the inner coma were easily discernible in unenhanced images. After 2016 January, the coma weakened, asymmetries became less obvious, and the dust trail became more pronounced. Throughout our observations, the nucleus was not detectable as it was significantly fainter than the inner coma. Adopting the nucleus parameters used in \citet{snodgrass13}, the nucleus was $\sim$2 mag fainter than the coma in a 10$^4$~km radius aperture during our earliest and latest observations, and more than 6 mag fainter near perihelion.

\begin{figure*}
  \centering
  \includegraphics[width=175mm]{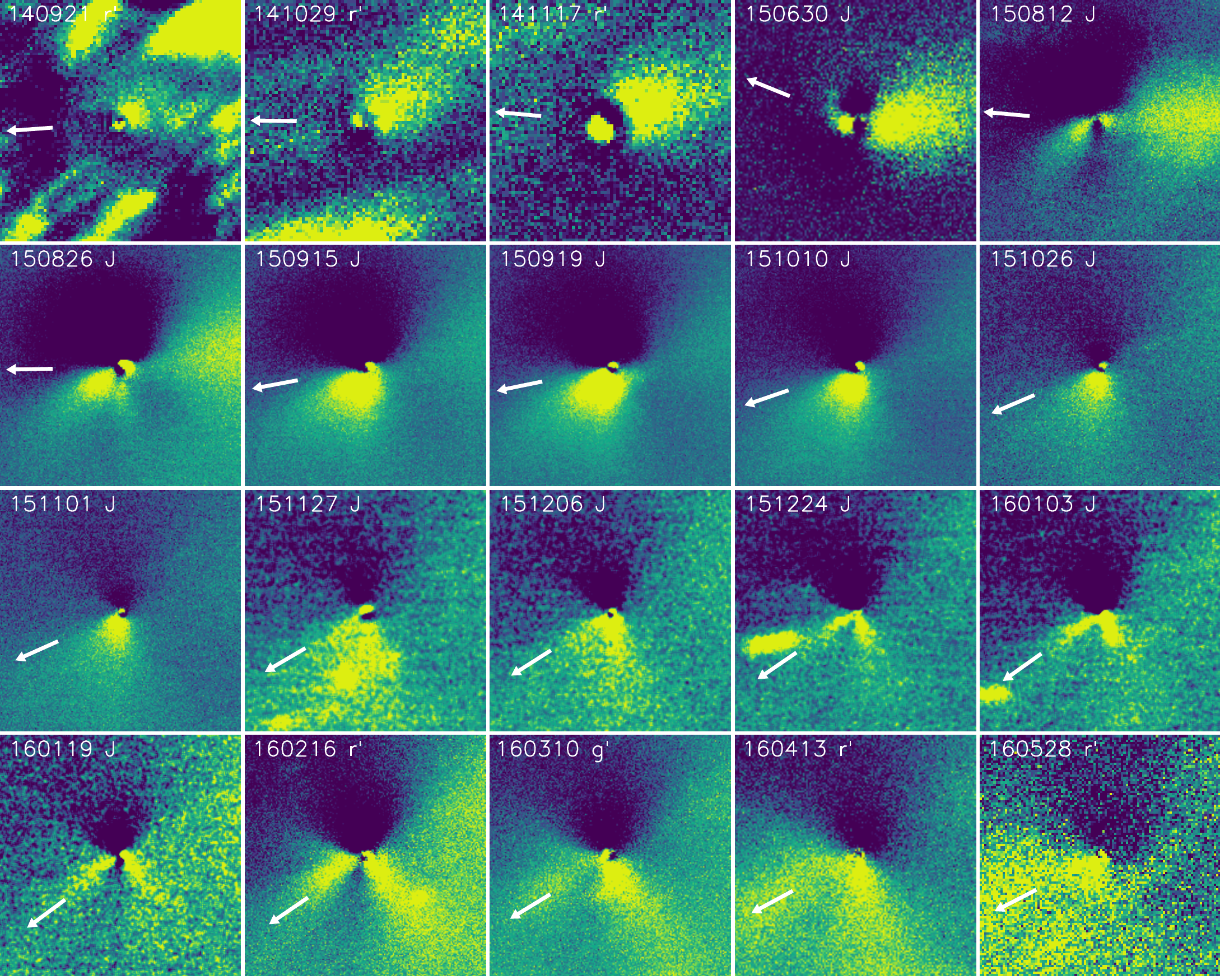}
  \caption[Composite of enhanced images]{Evolution of coma features from 2014 September through 2016 May for the same images as shown in Fig.~\ref{fig:dust_unenhanced}. All images have been enhanced by subtraction of an azimuthal median profile, are centered on the comet, and are 30,000~km across at the comet. Features within a few PSFs of the center should be interpreted with caution due to the enhancement process. Low signal-to-noise images have been smoothed with a boxcar smooth. All other details are as given in Fig.~\ref{fig:dust_unenhanced}.}
  \label{fig:dust_enhanced}
\end{figure*}

In order to investigate the faint, underlying structures in the inner coma, each nightly stacked image in a given filter was enhanced to remove the bulk coma brightness. We applied a variety of image enhancement techniques, as different methods are better at revealing different kinds of features \citep[e.g.,][]{schleicher04,samarasinha14}. In general, we found removal (division or subtraction) of an azimuthal median profile or a ${\rho}^{-1}$ profile (where $\rho$ is the projected distance from the nucleus) to be most effective. Before accepting a particular feature as real, we verified that it was discernible in the unenhanced images and, whenever signal-to-noise permitted, in the individual, unstacked images as well. Fig.~\ref{fig:dust_enhanced} shows images enhanced by subtracting an azimuthal median profile at representative times throughout the orbit. With this enhancement, the fainter features seen in the unenhanced images become prominent.

The enhancements did not reveal any obvious features other than the dust tail in the 2014 data. However, we note that the large geocentric distance and overall faintness of the comet would limit the angular extent of any features and thus make their detection challenging. This is particularly important for our preferred enhancement techniques, where features within a few point spread functions (PSFs) of the center are highly sensitive to centroiding effects and should be interpreted with caution. A faint, sunward feature was first detected in 2015 June, when it curved towards the north, presumably due to the effects of radiation pressure. When we next observed 67P in 2015 August, a distinct, straight sunward feature was seen at a position angle (P.A., measured counterclockwise from north through east) of $\sim$120$^\circ$. From 2015 September through December the sunward feature became wider and appeared to consist of two overlapping smaller features. From late December onwards these two features separated entirely, eventually having P.A.'s ${\sim}90^\circ$ apart, with the more southern feature essentially orthogonal to the sunward direction. From 2015 August through the end of our observations, the features extend projected distances at least $\sim$10$^4$~km from the nucleus. An additional, much fainter feature can be discerned at P.A.$\sim$80$^\circ$ in 2016 February--March, and possibly from 2015 December through 2016 April.  

We did not detect variability in the morphology during a night, but this is unsurprising given that all observations were snapshots lasting an hour or less. Assuming that a feature would need to have moved by at least a few times the seeing disc to be detectable, e.g., at least 2--3 arcsec, and that our typical observations were acquired at geocentric distances ($\mathit{\Delta}$) of 1.5--2.0 AU, the dust would have needed to move a projected distance of 2000--4000 km in an hour or less. This would require projected dust velocities on the order of 1~km~s$^{-1}$, which are unrealistic given the velocities reported by previous authors \citep[$\lesssim$35~m~s$^{-1}$; e.g.,][]{tozzi11}. 

We also did not detect significant changes in morphology from night to night. This is long relative to the nucleus' rotation period \citep[12.4~hr;][]{sierks15}. If the coma morphology was changing appreciably with rotation, we should have seen evidence for this from night to night, even allowing for the relative commensurability of rotational phase over consecutive nights since observations a few days apart would differ by about a quarter of a rotation.  Instead, the only significant variations in appearance occurred gradually during the apparition. This slow evolution of appearance is consistent with activity that is responding to the changing seasonal illumination of the nucleus rather than to diurnal variations in local illumination. A similar conclusion has been reached by other authors who imaged these features \citep[e.g.,][]{vincent13,boehnhardt16,zaprudin17,rosenbush17} and we will revisit this in Section~\ref{sec:disc}.

At speeds of a few tens of meters per second, the time for dust grains to completely traverse a feature will be of order a few days. Thus the features contain material ejected over multiple rotation periods, making it challenging to detect rotational variation. If the grains have significant velocity dispersion then a potential rotational signal would be further suppressed, and even considerably better image resolution might not reveal variability. 

\begin{figure*}
  \centering
  \includegraphics[width=150mm]{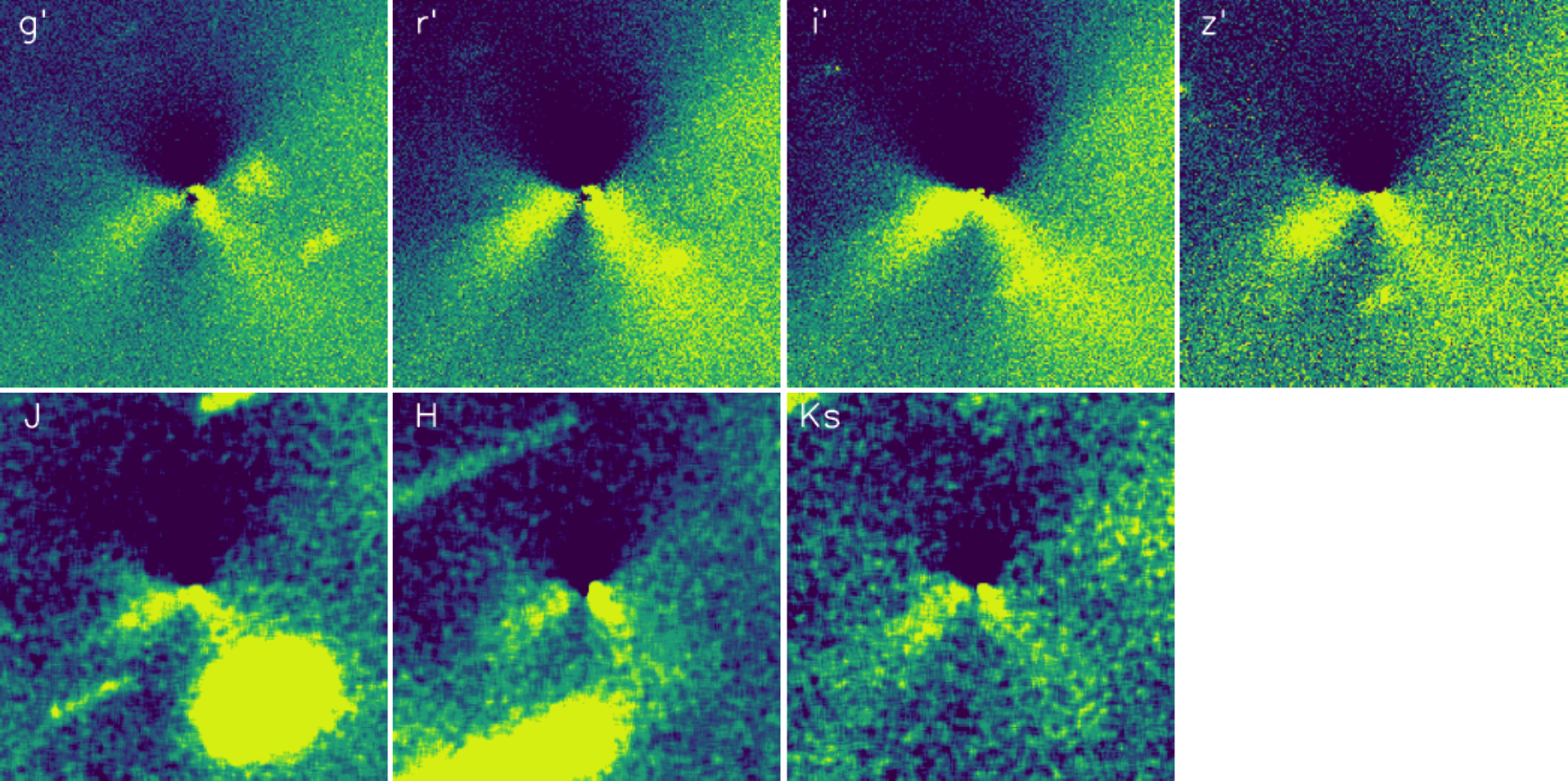}
  \caption[Composite of enhanced images on the same date]{Multi-filter coma morphology on 2016 February 16 showing similar morphology in all filters. The filter is given on each image. All other image details are the same as in Fig.~\ref{fig:dust_enhanced}.}
  \label{fig:all_filters}
\end{figure*}

We acquired images in seven bandpasses ($g'$, $r'$, $i'$, $z'$, $J$, $H$, and $Ks$) on one night, 2016 February 16. Fig.~\ref{fig:all_filters} shows enhanced versions of each. The same features are seen with approximately the same extent and relative brightness ($\sim$30\% of the coma brightness at that projected distance) in all seven panels. This supports our assumption that there was not substantial gas contamination in $g'$, and we interpret the similarity as indicating that there are not significant differences in the particles that dominate the scattering cross section in each bandpass. Ratios of unenhanced images (e.g., $g'/r'$, $J/r'$) did not show any obvious spatial differences that might be explained by grain properties. However, we note that such analyses are made challenging by the presence of faint stars near the nucleus in all visible-light frames and the very bright star near the nucleus in the $J$ and $H$ frames.

\subsubsection{Gas Morphology}
\begin{figure}
  \centering
  \includegraphics[width=85mm]{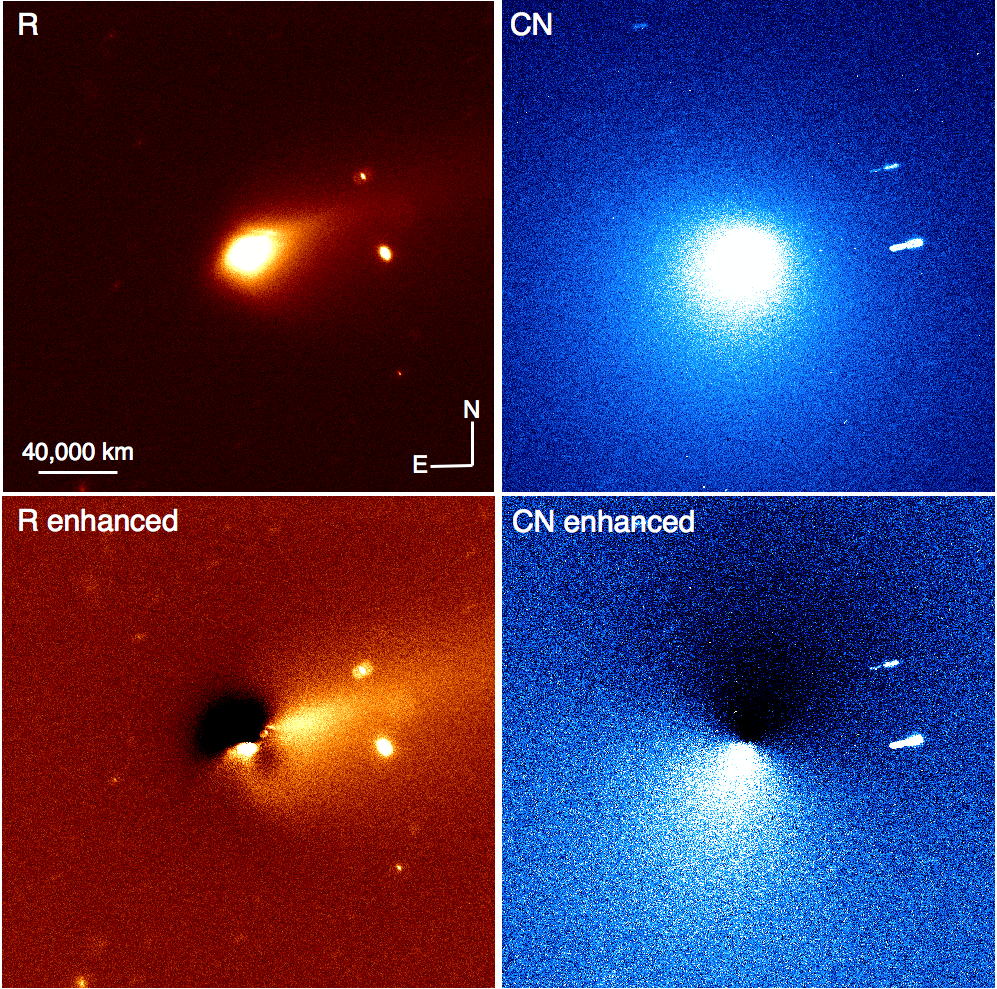}  
  \caption[Dust vs gas morphology]{Comparison of dust ($R$-band, left column) and gas ({\it CN}, right column) morphology on 2015 September 23. Unenhanced images are on the top row, and images enhanced by subtraction of an azimuthal median profile are on the bottom row. The $R$ images show the tail to the northwest (P.A.$\sim$285$^\circ$) and two shorter, sunward-facing features; these features are absent in the {\it CN} images, which show a bulk enhancement of brightness in the sunward hemisphere, but no distinct, small-scale features. All images are centered on the nucleus, have north up and east to the left, and have the same scale. Stars are seen trailed roughly east-west.}
  \label{fig:dustvsgas}
\end{figure}

Our data collection was primarily focussed on filters dominated by dust, but images were also acquired with the HB narrowband {\it CN} filter \citep{farnham00} on 22 nights when 67P was brightest. The {\it CN} bandpass contains both emission from the bright CN bandhead at 3883~\AA\ and reflected solar continuum from the dust. Ideally, we would remove the underlying solar continuum to produce a pure CN gas image, however, this was not possible on most nights due to observing in non-photometric conditions. Fortunately, comparison of both the raw and enhanced images (Fig.~\ref{fig:dustvsgas}) reveals that the coma morphology is very different between the $R$ and {\it CN} filters, indicating that dust contamination is not significant -- even the prominent dust tail is not seen appreciably in {\it CN} images. Thus, we can safely assess the CN morphology without needing to decontaminate the images.

Fig.~\ref{fig:dustvsgas} shows $R$ and {\it CN} images enhanced by subtraction of an azimuthal median profile on 2015 September 23, the highest signal-to-noise {\it CN} image we acquired. $R$ exhibits a similar morphology to contemporaneous near-IR images shown in Fig.~\ref{fig:dust_enhanced}, with two features in the sunward direction at P.A.'s$\sim$130$^\circ$ and 175$^\circ$ and the dust tail to the west in the anti-solar direction. The {\it CN} image does not show any of these features. Instead, it is visibly asymmetric even in the raw images. Enhancement reveals a hemispheric increase in brightness to the southeast with the brightest region near P.A.$\sim$165$^\circ$ and brightness decreasing roughly symmetrically at larger and smaller P.A.'s. Although high signal-to-noise {\it CN} images were only obtained on three nights with DCT, the much lower signal-to-noise {\it CN} images from the 31-in also show this hemispheric enhancement to the south. The P.A. of the peak brightness is approximately coincident with the brightest dust feature, suggesting that the CN originates from the same source region(s) as the dust.

Owing to 67P's orientation, the projection to the south in our images corresponds to the region above the nucleus' southern hemisphere as defined by {\it Rosetta}. Thus, our observations suggest that CN originates from the nucleus' southern hemisphere, in agreement with the TRAPPIST results that show a strong seasonal dependence when the nucleus' southern hemisphere is illuminated \citep{opitom17}. This is unlike the behavior of water and dust, which generally follow the heliocentric distance \citep[e.g.,][]{fougere16,hansen16}, and may suggest the parent of CN is not tied to the water production. ROSINA measurements reported by \citet{luspaykuti15} showed that HCN generally follows the water production, so if CN is not coupled to the water then HCN may not be its primary parent. However, \citet{luspaykuti15} looked at rotational timescales; the long-term evolution has not yet been demonstrated.

We do not see evidence of smaller structures (e.g., jets) in the CN coma in any of our data, nor do we see significant variations from night to night. CN coma structures were first discovered in 1P/Halley \citep{ahearn86} and are frequently seen in other comets. They are commonly used as a means by which to constrain the rotation period, pole orientation, and/or location and number of active regions on the surface \citep[e.g., ][]{farnham07a,knight11b}. The lack of obvious structures in 67P's CN coma as well as the hemispheric nature may simply be due to the low signal-to-noise. These structures are typically $\lesssim$10\% brighter than the ambient coma at the same projected distance, and 67P was fainter than most comets for which we have detected such structures. Furthermore, $\mathit{\Delta}$ was reasonably large ($>$1.7~AU), limiting the angular extent of any structures. 

If the lack of CN coma structures is real, it may suggest that the activity producing the parent(s) of CN is widespread and not confined to a particular, isolated active region. This is consistent with the overall picture of activity thus far revealed by {\it Rosetta}, where much of the sunlit surface seems to be active at low levels \citep[e.g.,][]{fougere16,hansen16}. Alternatively, the hemispheric distribution of CN could be due to its originating from grains in the coma rather than from one or more parent species leaving the nucleus directly as a gas. If significant amounts of CN comes from grains in the coma (a so called ``extended source''), this would naturally suppress the rotational variation that is typically a hallmark of CN in comet comae. The excess velocity from vapourization of grains would be in random directions and would greatly exceed the velocity of the grains from which the CN originated, and the grains themselves would likely have lost the rotational signature due to their presumed velocity dispersion. Furthermore, grains in the coma would be illuminated equally well in all parts of the coma. Thus, CN could originate from anywhere in the coma, and the relative enhancement in the sunward hemisphere would simply be due to there being more fresh grains in that direction (as dust grains in the tail would be older and more likely to have already lost their CN).

\subsection{Activity}
A common method for assessing cometary activity is via the $Af{\rho}$ parameter \citep{ahearn84}, which is often used as a proxy for dust production. Here $A$ is the phase angle ($\theta$) dependent albedo of dust in the aperture, $f$ is the filling factor of the dust, and $\rho$ is the projected aperture distance at the comet. We measured $A({\theta})f{\rho}$ for all visible-light data in a ${\rho}=10^4$~km radius aperture except for the 2014 Gemini data where the fields were too crowded, necessitating using a ${\rho}=5{\times}10^3$~km aperture. An aperture of ${\rho}=10^4$~km is common for $Af{\rho}$ measurements, allowing us to compare our data with those of other observers. Only images without obvious stars in the aperture were used and data were adjusted to 0$^\circ$ phase angle using the Halley-Marcus dust phase function\footnote{\href{http://asteroid.lowell.edu/comet/dustphase.html}{http://asteroid.lowell.edu/comet/dustphase.html}} \citep{schleicher98,marcus07b}, yielding $A(0^{\circ})f{\rho}$. The phase function does not have an analytical fit, but the interpolated correction for each night is given in column 3 of Table~\ref{t:mag}.

\begin{table*}
	\centering
	\setlength{\tabcolsep}{0.04in}
	\caption{Apparent magnitude and phase angle corrected $Af{\rho}$ for visible-light data}
	\label{t:mag}
	\begin{tabular}{lcccccccccccccccccc}
		\hline
  UT Date&Tel.$^{a}$&$\rho$$^{b}$&Phase&\multicolumn{10}{c}{Apparent magnitude}&\multicolumn{5}{c}{$A(0^\circ)f{\rho}$ (cm)}\\
  \cmidrule(lr){5-14}
  \cmidrule(lr){15-19}
  &&(km)&corr.$^{c}$&$m_R$&${\sigma}_{m_R}$&$m_{g'}$&${\sigma}_{m_{g'}}$&$m_{r'}$&${\sigma}_{m_{r'}}$&$m_{i'}$&${\sigma}_{m_{i'}}$&$m_{z'}$&${\sigma}_{m_{z'}}$&$R$&$g'$&$r'$&$i'$&$z'$\\  

		\hline
2014 Sep 20&GS&$5{\times}10^3$&-0.649&-&-&21.017&-&20.222&0.042&20.011&-&20.152&-&-&73&78&69&52\\
2014 Oct 29&GS&$5{\times}10^3$&-0.652&-&-&20.738&-&19.984&0.045&-&-&-&-&-&99&102&-&-\\
2014 Nov 11&GS&$5{\times}10^3$&-0.619&-&-&-&-&19.981&0.067&-&-&-&-&-&-&99&-&-\\
2014 Nov 12&GS&$5{\times}10^3$&-0.616&-&-&-&-&19.953&0.034&-&-&-&-&-&-&101&-&-\\
2014 Nov 13&GS&$5{\times}10^3$&-0.613&-&-&-&-&19.943&0.022&-&-&-&-&-&-&101&-&-\\
2014 Nov 15&GS&$5{\times}10^3$&-0.606&-&-&20.566&-&19.896&-&19.626&-&19.442&-&-&110&105&98&101\\
2014 Nov 16&GS&$5{\times}10^3$&-0.603&-&-&-&-&19.976&0.032&-&-&-&-&-&-&97&-&-\\
2014 Nov 17&GS&$5{\times}10^3$&-0.600&-&-&-&-&19.908&0.027&-&-&-&-&-&-&103&-&-\\
2014 Nov 18&GS&$5{\times}10^3$&-0.594&-&-&-&-&19.852&0.021&-&-&-&-&-&-&108&-&-\\
2014 Nov 19&GS&$5{\times}10^3$&-0.591&-&-&20.885&-&19.462&0.038&19.608&-&19.462&-&-&80&153&98&97\\
2015 Aug 18&31in&10$^4$&-1.023&13.420&0.030&-&-&-&-&-&-&-&-&1366&-&-&-&-\\
2015 Aug 19&31in&10$^4$&-1.024&13.386&0.039&-&-&-&-&-&-&-&-&1411&-&-&-&-\\
2015 Aug 20&31in&10$^4$&-1.024&13.366&0.040&-&-&-&-&-&-&-&-&1440&-&-&-&-\\
2015 Aug 21&31in&10$^4$&-1.024&13.345&0.059&-&-&-&-&-&-&-&-&1470&-&-&-&-\\
2015 Aug 22&31in&10$^4$&-1.024&13.227&0.025&-&-&-&-&-&-&-&-&1642&-&-&-&-\\
2015 Aug 23&31in&10$^4$&-1.024&13.283&0.023&-&-&-&-&-&-&-&-&1562&-&-&-&-\\
2015 Aug 29&31in&10$^4$&-1.023&13.243&0.034&-&-&-&-&-&-&-&-&1649&-&-&-&-\\
2015 Sep 12&31in&10$^4$&-1.021&13.317&0.058&-&-&-&-&-&-&-&-&1650&-&-&-&-\\
2015 Sep 18&31in&10$^4$&-1.020&13.399&0.010&-&-&-&-&-&-&-&-&1594&-&-&-&-\\
2015 Sep 19&31in&10$^4$&-1.020&13.312&0.038&-&-&-&-&-&-&-&-&1739&-&-&-&-\\
2015 Sep 20&31in&10$^4$&-1.020&13.426&0.033&-&-&-&-&-&-&-&-&1577&-&-&-&-\\
2015 Sep 23&DCT&10$^4$&-1.018&13.548&0.054&-&-&-&-&-&-&-&-&1439&-&-&-&-\\
2015 Sep 24&DCT&10$^4$&-1.018&13.580&0.034&-&-&-&-&-&-&-&-&1409&-&-&-&-\\
2015 Sep 24&31in&10$^4$&-1.018&13.595&0.036&-&-&-&-&-&-&-&-&1390&-&-&-&-\\
2015 Sep 25&31in&10$^4$&-1.018&13.633&0.013&-&-&-&-&-&-&-&-&1354&-&-&-&-\\
2015 Sep 26&31in&10$^4$&-1.018&13.647&0.013&-&-&-&-&-&-&-&-&1346&-&-&-&-\\
2015 Sep 30&31in&10$^4$&-1.018&13.763&0.080&-&-&-&-&-&-&-&-&1251&-&-&-&-\\
2015 Oct 2&31in&10$^4$&-1.017&13.809&0.029&-&-&-&-&-&-&-&-&1218&-&-&-&-\\
2015 Oct 3&31in&10$^4$&-1.017&13.820&0.021&-&-&-&-&-&-&-&-&1218&-&-&-&-\\
2015 Oct 13&31in&10$^4$&-1.015&14.170&0.059&-&-&-&-&-&-&-&-&959&-&-&-&-\\
2015 Nov 2&31in&10$^4$&-1.013&14.652&0.023&-&-&-&-&-&-&-&-&726&-&-&-&-\\
2015 Nov 8&31in&10$^4$&-1.012&14.822&0.030&-&-&-&-&-&-&-&-&649&-&-&-&-\\
2015 Nov 9&31in&10$^4$&-1.012&14.898&0.025&-&-&-&-&-&-&-&-&609&-&-&-&-\\
2015 Nov 10&31in&10$^4$&-1.012&14.906&0.039&-&-&-&-&-&-&-&-&609&-&-&-&-\\
2015 Nov 18&31in&10$^4$&-1.009&15.075&0.026&-&-&-&-&-&-&-&-&548&-&-&-&-\\
2015 Nov 19&31in&10$^4$&-1.007&15.113&0.028&-&-&-&-&-&-&-&-&532&-&-&-&-\\
2015 Nov 20&31in&10$^4$&-1.007&15.108&0.039&-&-&-&-&-&-&-&-&538&-&-&-&-\\
2015 Nov 27&31in&10$^4$&-1.002&15.279&0.032&-&-&-&-&-&-&-&-&476&-&-&-&-\\
2015 Nov 30&31in&10$^4$&-1.001&15.371&0.023&-&-&-&-&-&-&-&-&443&-&-&-&-\\
2015 Dec 1&31in&10$^4$&-0.999&15.383&0.019&-&-&-&-&-&-&-&-&440&-&-&-&-\\
2015 Dec 6&DCT&10$^4$&-0.994&15.369&0.046&-&-&-&-&-&-&-&-&454&-&-&-&-\\
2015 Dec 7&DCT&10$^4$&-0.992&15.373&0.086&-&-&-&-&-&-&-&-&453&-&-&-&-\\
2016 Feb 16&GN&10$^4$&-0.513&-&-&16.440&0.006&15.318&0.006&15.947&0.005&14.737&0.017&-&271&392&161&423\\
2016 Mar 10&GN&10$^4$&-0.169&-&-&16.480&0.012&-&-&-&-&-&-&-&242&-&-&-\\
2016 Apr 13&GN&10$^4$&-0.516&-&-&18.259&0.030&17.977&0.185&16.959&0.052&16.552&0.062&-&120&80&149&188\\
2016 Apr 30&GN&10$^4$&-0.631&-&-&18.572&0.014&18.060&0.010&17.526&0.029&16.886&0.075&-&142&118&141&219\\
2016 May 23&GN&10$^4$&-0.693&-&-&19.271&0.091&18.142&0.030&18.813&0.035&17.472&0.078&-&126&184&73&215\\
2016 May 28&GN&10$^4$&-0.696&-&-&-&-&18.906&0.060&-&-&-&-&-&-&101&-&-\\
		\hline
\multicolumn{19}{l}{$^{a}$ Telescope: GS = Gemini South, GN = Gemini North, DCT = Discovery Channel Telescope, 31in = Lowell 31-in.}\\		
\multicolumn{19}{l}{$^{b}$ Aperture radius}\\
\multicolumn{19}{l}{$^{c}$ Correction in magnitudes to 0$^\circ$ phase angle using the Halley-Marcus phase function  \citep{schleicher98,marcus07b}.}\\
	\end{tabular}
\end{table*}

\begin{figure*}
  \centering
  \includegraphics[width=140mm]{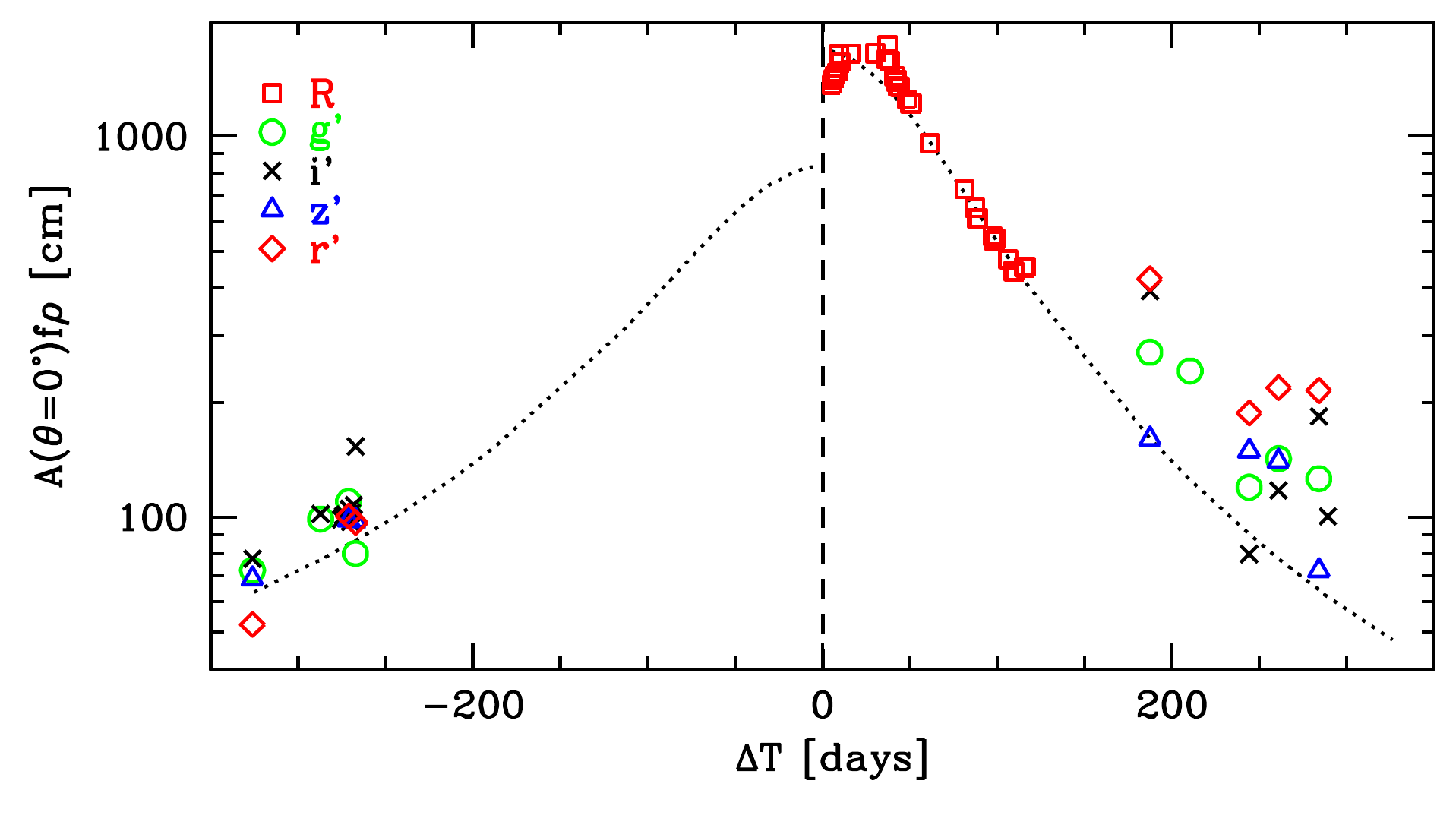}
  \caption[Visible $Af\rho$]{Visible-light $A({\theta}=0^{\circ})f{\rho}$ during the apparition. Uncertainties are not plotted, but can be calculated from Table~\ref{t:mag} and are generally smaller than the data points. The vertical dashed line denotes perihelion. The  dotted curve is the best-fit H$_2$O production rate \citep{hansen16} from all {\it Rosetta} instruments (pre-perihelion) and from ROSINA (post-perihelion) scaled to line up with our data. Note that this curve does not attempt to fit the post-perihelion shift in the peak water production, which \citet{hansen16} found to occur at $+$18 to $+$22 days.}
  \label{fig:afrho}
\end{figure*}

$A(0^{\circ})f{\rho}$ is plotted as a function of time relative to perihelion in Fig.~\ref{fig:afrho} and given along with apparent magnitudes in the same apertures in Table~\ref{t:mag}. Uncertainties are not plotted as they are generally smaller than the points, but the magnitude uncertainties are given in the table and can be propagated into $A(0^{\circ})f{\rho}$ uncertainties as ${\sigma}_{Af{\rho},X}=[1-10^{(-0.4{\times}{\pm}{\sigma}_{m_X})}]{\times}A(0^{\circ})f{\rho},_X$ where $X$ is the filter under consideration and the $\pm$ in the exponent provides lower and upper uncertainties. Magnitude uncertainties were calculated as the standard deviation of all the comet's individual magnitude measurements on a night added in quadrature to the uncertainty in the absolute calibrations. The absolute calibration uncertainty is only known for the Lowell data which were calibrated from field stars; default extinction correction values were used for the Gemini data and are estimated to be accurate to $\sim$5\%\footnote{\href{https://www.gemini.edu/sciops/instruments/gmos/calibration/baseline-calibs}{https://www.gemini.edu/sciops/instruments/gmos/calibration/{\newline}baseline-calibs}}, but they are not included in the magnitude uncertainty. When only one image was acquired no uncertainty is given, but it is likely 0.05--0.10~mag. We excluded a handful of nights in which conditions were clearly non-photometric. 

Throughout the apparition, the apparent $R$-band magnitudes agree well with the predictions from \citet{snodgrass13}, indicating that the overall activity remained relatively constant from 2009 to 2015. $A(0^{\circ})f{\rho}$ reaches a peak at about $+$30 days relative to perihelion ($\Delta$T) and then decreases smoothly. The timing of the peak is consistent with other ground-based data sets \citep[e.g.,][]{weiler04,schleicher06b,boehnhardt16,opitom17} that generally found a maximum in activity about one month post-perihelion. Fig.~\ref{fig:afrho} also plots the water production as measured by instruments on Rosetta \citep[e.g.,][]{hansen16}. $A(0^{\circ})f{\rho}$ follows this curve well except for the interval from 0 to $+$30 days when \citet{hansen16} did not attempt to account for the shift in peak water production which they reported occurs at 18--22 days post-perihelion. The agreement between these curves and our data support the conclusion that the dust production is tied to water production \citep[e.g.,][]{hansen16}.

At large $r_\mathrm{H}$, $A(0^{\circ})f{\rho}$ is consistently higher than the water production curve, likely due to several factors. First, larger and/or slower moving grains that were released earlier remain in the aperture. Second, the nucleus begins to contribute non-negligibly ($\sim$10\%) to the total signal. Third, the tail becomes increasingly foreshortened as viewed from Earth, contributing more signal to the aperture. 

Although the pre- and post-perihelion GMOS color data cannot be compared directly due to the differing aperture size, one night (2014 November 12) had a clear enough field to measure $A(0^{\circ})f{\rho}$ in a 10$^4$~km radius aperture. $A(0^{\circ})f{\rho}$ in the 10$^4$~km aperture was $\sim$40\% higher than in the $5{\times}10^3$~km aperture. If all of the 2014 data are scaled up by $\sim$40\%, it still appears that $A(0^{\circ})f{\rho}$ was higher post-perihelion, presumably due to the presence of large and/or low velocity grains released near perihelion remaining in the photometric aperture (grains released near the previous perihelion would have left the photometric aperture prior to our earliest observations). A bias for larger $A(0^{\circ})f{\rho}$ values at large heliocentric distances post-perihelion as well as the phenomenon of larger $Af{\rho}$ values being recorded in smaller apertures (implying a brightness profile steeper than ${\rho}^{-1}$) are common \citep[e.g.,][]{baum92}. 

From $\Delta\mathrm{T}=+$35 to $+$120 days (2015 September through December), $A(0^{\circ})f{\rho}$ data decreased as $r_\mathrm{H}^{-4.2}$. This is consistent with the findings of \citet{boehnhardt16} from 2015 September through 2016 May (power-law slope of $-$4.1 to $-$4.2). However, it is steeper than the slopes reported over comparable $r_\mathrm{H}$ ranges in previous apparitions by \citet{schleicher06b} ($-1.6{\pm}1.0$ for green continuum) and \citet{snodgrass13} ($-$3.4). The overall slope flattens significantly, to $-$2.6 for ${\Delta}\mathrm{T}>35$ days, if the GMOS $r'$ data are included and converted to $R$ values using the SDSS conversions\footnote{\href{http://classic.sdss.org/dr5/algorithms/sdssUBVRITransform.html}{http://classic.sdss.org/dr5/algorithms/sdssUBVRITransform.html}} and assuming solar color.   However, the data are not well fit by a single power-law over the full time interval because the conditions for comparing $A(0^{\circ})f{\rho}$ are not met at large $r_\mathrm{H}$ as previously discussed. The differences in power-law slopes between authors is likely due to differences in methodology (e.g., \citealt{snodgrass13} assume a linear phase angle correction, while \citealt{schleicher06b} use a non-constant aperture size and no phase angle correction), non-uniform sampling of our data (biasing towards a steeper fit), as well as intrinsic scatter in the data, which is significant due to our short integration times when the comet was faint.

\subsection{Possible Outbursts}
Table~\ref{t:mag} and Fig.~\ref{fig:afrho} reveal increases in brightness above the local trend and larger than the photometric uncertainties on 2015 August 22 ($+$9.3 days) and 2015 September 19 ($+$37.3 days) that appear to have been minor outbursts. These are not caused by intrusion of a passing star and do not appear to be due to miscalibration as field stars agree with their catalog magnitudes within $\pm$0.02~mag on August 22 and $\pm$0.04~mag on September 19. A linear fit to the magnitudes on August 18--21 predicts the comet should have been 0.09 mag fainter on August 22 than we measured, a 3-$\sigma$ result based on our estimated uncertainty. On August 23, the apparent magnitude was 0.01 mag brighter than the linear fit would predict, indicating that to within our photometric uncertainties the comet had returned to its normal brightening behavior. Similarly, a linear fit to September 18--30 but excluding September 19 predicts a magnitude 0.11 fainter on September 19 than was measured (also $\sim$3$\times$ the uncertainty for that night). September 20 was 0.03 mag brighter than the linear fit would predict, but this was within the photometric uncertainty of the linear fit, suggesting the comet had returned to its baseline behavior.  

The August 22 brightening appears to be the photometric signature of the outburst identified by \citet{boehnhardt16} on the basis of coma morphology in their 2015 August 23 data. \citet{boehnhardt16} did not see evidence of the outburst in their August 22 images or photometry, suggesting the outburst commenced some time between the end of their observations (3:17 UT) and the beginning of our observations at 11:17 UT. Inspection of our enhanced images does not show unusual coma morphology on August 22 or 23. For an outflow velocity of $\sim$150~m~s$^{-1}$ (as suggested by \citealt{boehnhardt16}), the outburst material would have traveled at most $\sim$4300~km before our first observation on August 22, which is roughly comparable to the seeing disk and would not be identifiable morphologically. By August 23, the ejected material had likely dispersed enough that it was not detectable in our 31-in images (effective aperture size of 0.7-m), which presumably had lower signal-to-noise than those acquired by \citet{boehnhardt16} with the 2-m telescope at Mt.\ Wendelstein. The return to the pre-outburst brightening rate on August 23 indicates that most of the outburst material was moving faster than the $\sim$115~m~s$^{-1}$ needed to have left the photometric aperture by this time.

We can see the signature of this outburst in our radial profiles, where the power-law slope of the coma brightness (${\rho}^{\alpha}$ from ${\rho}=4{\times}10^{3}-1.5{\times}10^{4}$~km) was approximately constant on August 20--21 (${\alpha}=-$1.77 and $-$1.79), but steeper on August 22 (${\alpha}=-$1.89) signifying additional material closer to the nucleus, and slightly flatter on August 23 (${\alpha}=-$1.75) signifying additional material further from the nucleus. These values are in good agreement with \citet{boehnhardt16}'s pre- and post-outburst slopes on August 22 and 23. (Note that for direct comparison between these results, +1 should be added to the slopes presented by \citealt{boehnhardt16} because they report the slope of $Af\rho$, which is proportional to flux/$\rho$, while we measured the slope of the flux.) After August 23 our next visible-light data were on August 29, at which time the slope was even flatter (${\alpha}=-$1.71). Although this flatter slope could be due to very slow material from the outburst moving outwards, the temporal gap from our earlier data makes it impossible to draw a firm conclusion. Our near-IR data were sampled too infrequently at this time (August 20 and 26) to exhibit a clear signature of the outburst.

We cannot look for the signature of the outburst with {\it CN} photometry because standard stars were not observed on any of these nights and there is no catalog available to calibrate the images with background stars. The slope of the CN radial profile does not show evidence of an outburst on August 22. This could be real and indicating that the outburst did not contain significant amounts of CN.  However, this may simply be due to the combination of the poor signal-to-noise of the {\it CN} images and large PSF (typical effective seeing for this telescope is 3--5 arcsec), which combined to flatten the {\it CN} profiles and made it difficult to accurately centroid. 

\citet{vincent16b} compiled outbursts seen by {\it Rosetta} near perihelion. Their event \#16 was detected by the NavCam at 6:47 UT on August 22, precisely in the window between the Mt.\ Wendelstein observations and our own. While it is tempting to ascribe the outburst we detected to this event, we caution that the ``relative luminosity'' of event \#16, as given by \citet{vincent16b} (Column 10 of their Table 1), was among the lowest in the sample. Several ostensibly stronger outbursts occurred less than 24~hr before our imaging on other nights and were not obvious in our data (e.g., event \#12 at 17:21 on August 12, event \#23 at 10:10 on August 28). Since we detected an increase of $\sim$10\% in the amount of coma material within ${\rho}=10^4$~km on August 22, if event \#16 was the source, the NavCam snapshot must have recorded merely a very small percentage of the total outburst. Furthermore, \citet{vincent16b} report a latitude for the outburst of $-40^\circ$, while modeling by \citet{boehnhardt16} suggested a source near latitudes $+5^\circ$ to $+10^\circ$. Thus, the identification of event \#16 as the source of the outburst we observed on August 22 is far from certain.

We are not aware of any reports of an outburst on or around 2015 September 19 to corroborate our possible detection. The coma's power-law slope is not significantly different on September 19 than on neighboring days, and we do not see obvious morphological changes in either $R$ or {\it CN}. The near-IR observations were taken too infrequently around this time to confirm an outburst photometrically and they do not show clear morphological evidence. Thus, we consider this only a potential outburst.

\section{Discussion}
\label{sec:disc}
While coma structures have been studied in a dozen or so comets, the vast majority of these have been studied on only one apparition, either because the comets have long periods or because viewing geometry has only permitted investigation on a single apparition during the modern era. A few comets have had near polar jets identified on multiple apparitions, e.g., 2P/Encke \citep{sekanina91b}, and 10P/Tempel~2 \citep{knight12}, allowing the orientation of their rotation axes to be determined and the stability of their rotational poles to be tested. Very few comets with structures in the coma that are attributed to lower latitude source regions have been observed on multiple apparitions \citep[e.g.,][]{farnham09}. 

\citet{schleicher06b} determined a pole solution for 67P based on dust coma features observed in 1996, and \citet{weiler04} published constraints on the pole based on dust coma features observed in 2004. \citet{vincent13} utilized these as well as dust features seen in 2009 to develop the most comprehensive model of 67P's pole and locations of active regions prior to {\it Rosetta}'s arrival. Their modeling constrained the rotational pole orientation within $\pm$10$^\circ$ in both RA and Dec, and could reproduce the observed activity during the 2003 and 2009 apparitions with three active regions located at latitudes $-$45$^\circ$, 0$^\circ$, and $+$60$^\circ$.

\begin{figure}
  \centering
  \includegraphics[width=85mm]{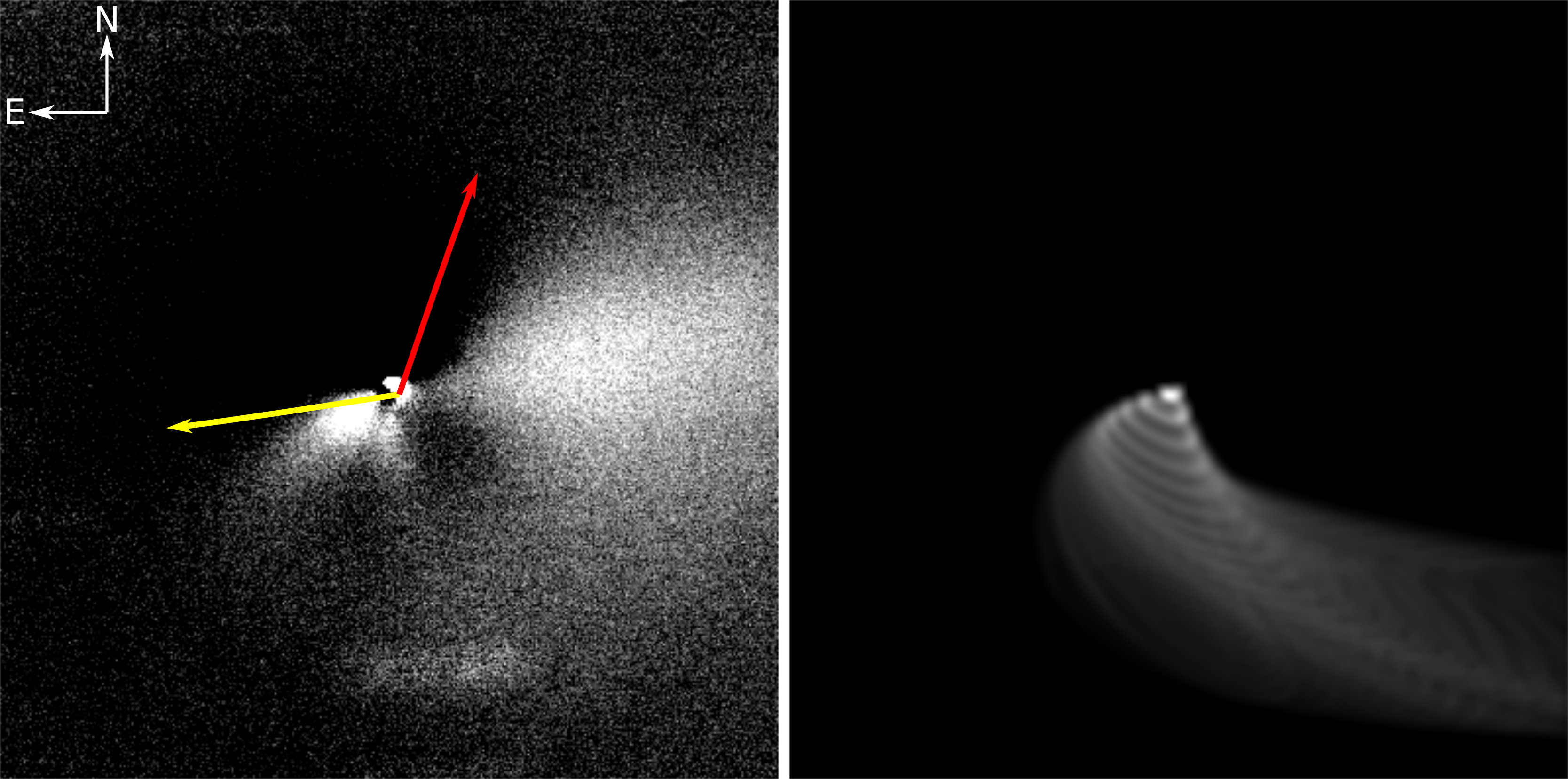}
  \caption[Image versus enhancement]{Enhanced $J$-band image from 2015 August 26 (left) compared with a simulated image from the same time using the model from \citet{vincent13}. The direction to the Sun is shown in yellow, the rotation axis is in red. North is up and east to the left. Each image is 60,000 km across. The model does not include the tail or bulk coma.
}
  \label{fig:model}
  \label{lastfig}
\end{figure}

We applied the \citet{vincent13} model without adjusting any parameters to our data, with the prediction for the morphology near perihelion shown in Fig.~\ref{fig:model}. The left panel shows our enhanced $J$-band image from 2015 August 26 while the right panel shows the model on the same scale. The model does not include the tail, which is in the opposite direction as the yellow arrow, at a P.A.$\sim$280$^\circ$. Overall, the modeled behavior replicates the observations this apparition reasonably well. The southern feature goes in the correct direction, but curves more than was observed. The sunward feature is in the predicted location but extends significantly farther in the data than in the simulations. Both of these features can be explained as being the edges of a single fan coming from the same source, where the optical depth through the edges is higher due to the projection of 3-D feature onto a 2-D image, thus causing it to be the only part evident in our enhanced images. The model can likely be improved by adjusting the dust size and velocity, as well as using the pole determined by {\it Rosetta}, but such investigations are beyond the scope of this paper. Our dataset's long temporal baseline with high spatial resolution provides excellent constraints for tracking the migration of activity across the surface during 67P's post-perihelion phase, and we hope to investigate this in detail in a subsequent paper.

Although we have only made a cursory investigation into the modeling at this time, the similarity of 67P's coma morphology to the predictions based on 2003 and 2009 is a compelling finding (we note that \citealt{boehnhardt16} and \citealt{zaprudin17} have also reached similar conclusions with their own datasets). This demonstrates that the comet's pole orientation and sources of activity have been relatively unchanged over at least the last three orbits. Despite the strong seasonal effects that see the nucleus' northern hemisphere covered with freshly deposited material during the brief but intense southern summer \citep[e.g.,][]{fornasier16}, the overall behavior is consistent and therefore the behavior observed by {\it Rosetta} is likely typical. With this knowledge, we can attempt to relate the active regions identified by \citet{vincent13} to specific terrain seen on the surface \citep{elmaarry15,elmaarry16}. Their high latitude ($+$60$^\circ$) region ``C'' likely corresponds with the Hapi region; the specific regions corresponding to their regions ``B'' ($-$45$^\circ$) and ``C'' (0$^\circ$) are less certain. Region ``B'' likely corresponds to the southern concavity while region ``A'' may correspond to Imhotep. Further work may allow us to tie specific properties of the coma features to their origin at the surface, providing a unique link between the large-scale Earth-based observations with the in situ observations made by {\it Rosetta}.

We note that a promising approach to bridging the gap between the Earth-based and in situ observations is provided by the modeling of \citet{kramer16} who replicated the general appearance of the many small jets seen within a few km of the surface by OSIRIS by simply modeling dust being released across the entire surface and responding to solar illumination. They found that the dust jets could be traced back to local concavities, both pits and craters as well as the large ``neck'' region between the two lobes of the nucleus. While \citet{kramer16} were not the first to model activity across the whole surface -- \citet{keller15} accurately predicted changes in the rotation period by modeling water sublimation -- or to propose this behavior -- specific linkages between jets and steep terrain had been identified previously by, e.g., \citet{farnham13} and \citet{vincent15,vincent16a}, while \citet{crifo02} argued that the morphology was affected by the shape of the comet -- their powerful modeling, if extended outward from the nucleus by 2--3 orders of magnitude, could demonstrate exactly how the numerous and variable small scale jets seen in the extreme inner coma produce the few large-scale, slowly varying features seen extending for thousands of km in our and other Earth-based images.

An important implication of the repeatable nature of 67P's activity from apparition to apparition is that remote observations in support of the {\it Rosetta} mission need not be confined to the 2015 apparition. As noted in the Introduction, 67P's 2015 apparition was very poor for Earth-based observers, making observations challenging and limiting the science that could be achieved from Earth. In contrast, the upcoming 2021 apparition will be 67P's best apparition since 1982, with $\mathit{\Delta}$ reaching a minimum of 0.42~AU and the comet visible in dark skies for more than a year around perihelion. These vastly superior observing conditions will permit far better observations than were achieved during the 2015 apparition, thus giving additional context to the {\it Rosetta} results.

Another upshot of this work is the value of regular observations of high photometric precision for detecting small outbursts in comets. Such outbursts have only been detected in a small number of comets \citep{ishiguro16} and their frequency across the comet population is as yet unknown. Monitoring of large numbers of comets for small outbursts will become possible in the near future with the Large Synoptic Survey Telescope (LSST). A quantitative assessment of the frequency and energy involved in outbursts may reveal if small outbursts are attributable to surface processes like cliff collapse or sub-surface thermal processes like amorphous-to-crystalline H$_2$O ice transitions.

\section{Summary}
\label{sec:summary}
We have presented our extensive observations of comet 67P/Churyumov-Gerasimenko using Gemini South and North and Lowell Observatory's DCT and 31-in telescopes in support of the {\it Rosetta} mission. We obtained usable data on 68 nights between 2014 September and 2016 May, with an effort made to obtain data regularly when the comet was visible from the ground and with a higher cadence at times of high interest to the mission ({\it Philae} landing and perihelion). We emphasized dust coma studies, as we obtained the most extensive $J$, $H$, $Ks$ dataset on 67P in existence and supplemented this with $g'$, $r'$, $i'$, $z'$ imaging early and late in the apparition, and $R$ and CN gas monitoring when the comet was brightest. 

Our major findings include:
\begin{enumerate}
\item We observed an evolution in 67P's coma morphology. From 2014 September through 2015 June it was dominated by a central condensation with the only significant feature being the tail. From 2015 August through 2016 May there were visible asymmetries in the inner coma. Upon enhancement these asymmetries were revealed to be due to one or more mostly straight features (i.e., ``jets''). These features spanned projected distances exceeding 10$^4$~km at the comet and were therefore much larger in scale than anything identified by the OSIRIS camera on {\it Rosetta}. Similar morphology was observed in all concurrently acquired broadband filters ($g'$, $r'$, $R$, $i'$, $z'$, $J$, $H$, $Ks$), which we interpret as indicating that the features were composed of dust grains having a range of particle sizes. Narrowband {\it CN} gas images had a very different appearance from the dust images, lacking distinct features and instead being brighter throughout the southern hemisphere of our images (corresponding to summer on the nucleus).

\item We measured dust production via the $Af{\rho}$ parameter and detected a clear peak $\sim$30 days after perihelion. Both the activity level and the time of peak activity are consistent with predictions from previous apparitions. $Af{\rho}$ decreased as $r_\mathrm{H}^{-4.2}$ from ${\Delta}\mathrm{T}=+$35 days to $+$120 days and as $r_\mathrm{H}^{-2.6}$ if the fading portion is extended to $+$300 days. The flattening slope at large heliocentric distances is likely due to large and slow moving grains remaining in the aperture and also causes the post-perihelion $Af{\rho}$ to be somewhat higher than the pre-perihelion value at comparable $r_\mathrm{H}$, but could also be due to other assumptions, e.g., the phase correction used.

\item We detected two apparent outbursts in our photometry, on 2015 August 22 and 2015 September 19. Neither was evident morphologically, and by the next night was not convincingly evident photometrically. When combined with observations by \citet{boehnhardt16}, we constrain the time of outburst on August 22 to 3:17--11:17 UT, and tentatively identify it with an event seen by {\it Rosetta} (\#16 in \citealt{vincent16a}), although the outburst we detected appears to have been significantly stronger than indicated by the snapshot acquired by {\it Rosetta}. We are not aware of any corroborating detections of the apparent outburst on September 19. We do not see clear evidence of any other outbursts despite several identified by \citet{vincent16a} that occurred within 24~hr preceding our observations.

\item The dust morphology can be replicated reasonably well using the model of \citet{vincent13} created from 2003 and 2009 observations. This indicates that the pole orientation and active areas on 67P have been relatively constant over at least three apparitions, and supports the conclusion that changes in 67P's morphology are primarily driven by the subsolar latitude.
\end{enumerate}

\newpage

\section*{Acknowledgements}
We thank Hermann Boehnhardt and Cyrielle Opitom for useful discussions regarding interpretation of our results and Andrew Stephens for help in processing data. We are grateful to the various Gemini staff for helping to plan and schedule our observations, and particularly thank Bruce Macintosh and the GPI Team for giving us a small part of their time to observe during the {\it Philae} landing phase. We also thank Hermann Boehnhardt for a careful and thorough review.

Based on observations obtained at the Gemini Observatory, which is operated by the Association of Universities for Research in Astronomy, Inc., under a cooperative agreement with the NSF on behalf of the Gemini partnership: the National Science Foundation (United States), the National Research Council (Canada), CONICYT (Chile), Ministerio de Ciencia, Tecnolog\'{i}a e Innovaci\'{o}n Productiva (Argentina), and Minist\'{e}rio da Ci\^{e}ncia, Tecnologia e Inova\c{c}\~{a}o (Brazil). Data were obtained under Gemini programs GS-2014B-Q-15-11, GS-2014B-Q-76, GS-2015A-Q-54, GN-2015B-Q-53, and GN-2016A-Q-53.

This work made use of the Discovery Channel Telescope at Lowell Observatory. Lowell is a private, non-profit institution dedicated to astrophysical research and public appreciation of astronomy and operates the DCT in partnership with Boston University, the University of Maryland, the University of Toledo, Northern Arizona University and Yale University. The Large Monolithic Imager was built by Lowell Observatory using funds provided by the National Science Foundation (AST-1005313).

This research has made use of the Gemini IRAF package as part of Ureka, which was provided by Space Telescope Science Institute and Gemini Observatory. It also used SAOImage DS9, developed by Smithsonian Astrophysical Observatory and the ``Aladin sky atlas'' developed at CDS, Strasbourg Observatory, France \citep{bonnarel00}.

M.M.K. and D.G.S. were supported by NASA's Planetary Astronomy grant NNX14AG81G. M.M.K. is grateful for office space provided by Johns Hopkins University Applied Physics Laboratory while working on this project. C.S. was supported by the UK STFC through an Ernest Rutherford Fellowship. B.C.C. acknowledges the support of the Australian Research Council through Discovery project DP150100862.




\bibliographystyle{mnras}

\begin{thebibliography}{}
\makeatletter
\relax
\def\mn@urlcharsother{\let\do\@makeother \do\$\do\&\do\#\do\^\do\_\do\%\do\~}
\def\mn@doi{\begingroup\mn@urlcharsother \@ifnextchar [ {\mn@doi@}
  {\mn@doi@[]}}
\def\mn@doi@[#1]#2{\def\@tempa{#1}\ifx\@tempa\@empty \href
  {http://dx.doi.org/#2} {doi:#2}\else \href {http://dx.doi.org/#2} {#1}\fi
  \endgroup}
\def\mn@eprint#1#2{\mn@eprint@#1:#2::\@nil}
\def\mn@eprint@arXiv#1{\href {http://arxiv.org/abs/#1} {{\tt arXiv:#1}}}
\def\mn@eprint@dblp#1{\href {http://dblp.uni-trier.de/rec/bibtex/#1.xml}
  {dblp:#1}}
\def\mn@eprint@#1:#2:#3:#4\@nil{\def\@tempa {#1}\def\@tempb {#2}\def\@tempc
  {#3}\ifx \@tempc \@empty \let \@tempc \@tempb \let \@tempb \@tempa \fi \ifx
  \@tempb \@empty \def\@tempb {arXiv}\fi \@ifundefined
  {mn@eprint@\@tempb}{\@tempb:\@tempc}{\expandafter \expandafter \csname
  mn@eprint@\@tempb\endcsname \expandafter{\@tempc}}}

\bibitem[\protect\citeauthoryear{{A'Hearn}, {Schleicher}, {Millis}, {Feldman}
  \& {Thompson}}{{A'Hearn} et~al.}{1984}]{ahearn84}
{A'Hearn} M.~F.,  {Schleicher} D.~G.,  {Millis} R.~L.,  {Feldman} P.~D.,
  {Thompson} D.~T.,  1984, \mn@doi [\aj] {10.1086/113552}, \href
  {http://adsabs.harvard.edu/abs/1984AJ.....89..579A} {89, 579}

\bibitem[\protect\citeauthoryear{{A'Hearn}, {Hoban}, {Birch}, {Bowers},
  {Martin}  \& {Klinglesmith}}{{A'Hearn} et~al.}{1986}]{ahearn86}
{A'Hearn} M.~F.,  {Hoban} S.,  {Birch} P.~V.,  {Bowers} C.,  {Martin} R.,
  {Klinglesmith} III D.~A.,  1986, \mn@doi [\nat] {10.1038/324649a0}, \href
  {http://adsabs.harvard.edu/abs/1986Natur.324..649A} {324, 649}

\bibitem[\protect\citeauthoryear{{Baum}, {Kreidl}  \& {Schleicher}}{{Baum}
  et~al.}{1992}]{baum92}
{Baum} W.~A.,  {Kreidl} T.~J.,   {Schleicher} D.~G.,  1992, \mn@doi [\aj]
  {10.1086/116310}, \href {http://adsabs.harvard.edu/abs/1992AJ....104.1216B}
  {104, 1216}

\bibitem[\protect\citeauthoryear{{Boehnhardt}, {Riffeser}, {Kluge}, {Ries},
  {Schmidt}  \& {Hopp}}{{Boehnhardt} et~al.}{2016}]{boehnhardt16}
{Boehnhardt} H.,  {Riffeser} A.,  {Kluge} M.,  {Ries} C.,  {Schmidt} M.,
  {Hopp} U.,  2016, \mn@doi [\mnras] {10.1093/mnras/stw2859}, \href
  {http://adsabs.harvard.edu/abs/2016MNRAS.462S.376B} {462, S376}

\bibitem[\protect\citeauthoryear{{Bonnarel} et~al.,}{{Bonnarel}
  et~al.}{2000}]{bonnarel00}
{Bonnarel} F.,  et~al., 2000, \mn@doi [\aaps] {10.1051/aas:2000331}, \href
  {http://cdsads.u-strasbg.fr/abs/2000A%26AS..143...33B} {143, 33}

\bibitem[\protect\citeauthoryear{{Bramich}}{{Bramich}}{2008}]{bramich08}
{Bramich} D.~M.,  2008, \mn@doi [\mnras] {10.1111/j.1745-3933.2008.00464.x},
  \href {http://esoads.eso.org/abs/2008MNRAS.386L..77B} {386, L77}

\bibitem[\protect\citeauthoryear{{Crifo}, {Rodionov}, {Szeg{\"o}}  \&
  {Fulle}}{{Crifo} et~al.}{2002}]{crifo02}
{Crifo} J.-F.,  {Rodionov} A.~V.,  {Szeg{\"o}} K.,   {Fulle} M.,  2002, Earth
  Moon and Planets, \href
  {http://adsabs.harvard.edu/abs/2002EM\%26P...90..227C} {90, 227}

\bibitem[\protect\citeauthoryear{{Eikenberry} et~al.,}{{Eikenberry}
  et~al.}{2004}]{eikenberry04}
{Eikenberry} S.~S.,  et~al., 2004, in {Moorwood} A.~F.~M.,  {Iye} M.,  eds,
  \procspie Vol. 5492, Ground-based Instrumentation for Astronomy. pp
  1196--1207, \mn@doi{10.1117/12.549796}

\bibitem[\protect\citeauthoryear{{El-Maarry} et~al.,}{{El-Maarry}
  et~al.}{2015}]{elmaarry15}
{El-Maarry} M.~R.,  et~al., 2015, \mn@doi [\aap] {10.1051/0004-6361/201525723},
  \href {http://adsabs.harvard.edu/abs/2015A%26A...583A..26E} {583, A26}

\bibitem[\protect\citeauthoryear{{El-Maarry} et~al.,}{{El-Maarry}
  et~al.}{2016}]{elmaarry16}
{El-Maarry} M.~R.,  et~al., 2016, \mn@doi [\aap] {10.1051/0004-6361/201628634},
  \href {http://adsabs.harvard.edu/abs/2016A%26A...593A.110E} {593, A110}

\bibitem[\protect\citeauthoryear{{Farnham}}{{Farnham}}{2009}]{farnham09}
{Farnham} T.~L.,  2009, \mn@doi [\planss] {10.1016/j.pss.2009.02.008}, \href
  {http://adsabs.harvard.edu/abs/2009P\%26SS...57.1192F} {57, 1192}

\bibitem[\protect\citeauthoryear{{Farnham}, {Schleicher}  \&
  {A'Hearn}}{{Farnham} et~al.}{2000}]{farnham00}
{Farnham} T.~L.,  {Schleicher} D.~G.,   {A'Hearn} M.~F.,  2000, \mn@doi
  [Icarus] {10.1006/icar.2000.6420}, \href
  {http://adsabs.harvard.edu/abs/2000Icar..147..180F} {147, 180}

\bibitem[\protect\citeauthoryear{{Farnham}, {Samarasinha}, {Mueller}  \&
  {Knight}}{{Farnham} et~al.}{2007}]{farnham07a}
{Farnham} T.~L.,  {Samarasinha} N.~H.,  {Mueller} B.~E.~A.,   {Knight} M.~M.,
  2007, \mn@doi [\aj] {10.1086/513186}, \href
  {http://adsabs.harvard.edu/abs/2007AJ....133.2001F} {133, 2001}

\bibitem[\protect\citeauthoryear{{Farnham}, {Bodewits}, {Li}, {Veverka},
  {Thomas}  \& {Belton}}{{Farnham} et~al.}{2013}]{farnham13}
{Farnham} T.~L.,  {Bodewits} D.,  {Li} J.-Y.,  {Veverka} J.,  {Thomas} P.,
  {Belton} M.~J.~S.,  2013, \mn@doi [\icarus] {10.1016/j.icarus.2012.06.019},
  \href {http://adsabs.harvard.edu/abs/2013Icar..222..540F} {222, 540}

\bibitem[\protect\citeauthoryear{{Feldman}, {Cochran}  \& {Combi}}{{Feldman}
  et~al.}{2004}]{feldman04}
{Feldman} P.~D.,  {Cochran} A.~L.,   {Combi} M.~R.,  2004, in {Festou, M.~C.,
  Keller, H.~U., \& Weaver, H.~A.} ed., , Comets II.
University of Arizona Press, Tucson, pp 425--447

\bibitem[\protect\citeauthoryear{{Fornasier} et~al.,}{{Fornasier}
  et~al.}{2016}]{fornasier16}
{Fornasier} S.,  et~al., 2016, \mn@doi [Science] {10.1126/science.aag2671},
  \href {http://adsabs.harvard.edu/abs/2016Sci...354.1566F} {354, 1566}

\bibitem[\protect\citeauthoryear{{Fougere} et~al.,}{{Fougere}
  et~al.}{2016}]{fougere16}
{Fougere} N.,  et~al., 2016, \mn@doi [\aap] {10.1051/0004-6361/201527889},
  \href {http://adsabs.harvard.edu/abs/2016A%26A...588A.134F} {588, A134}

\bibitem[\protect\citeauthoryear{{Gimeno} et~al.,}{{Gimeno}
  et~al.}{2016}]{gimeno16}
{Gimeno} G.,  et~al., 2016, in Society of Photo-Optical Instrumentation
  Engineers (SPIE) Conference Series. p. 99082S, \mn@doi{10.1117/12.2233883}

\bibitem[\protect\citeauthoryear{{Hansen} et~al.,}{{Hansen}
  et~al.}{2016}]{hansen16}
{Hansen} K.~C.,  et~al., 2016, \mn@doi [\mnras] {10.1093/mnras/stw2413}, \href
  {http://adsabs.harvard.edu/abs/2016MNRAS.462S.491H} {462, S491}

\bibitem[\protect\citeauthoryear{{Hodapp} et~al.,}{{Hodapp}
  et~al.}{2003}]{hodapp03}
{Hodapp} K.~W.,  et~al., 2003, \mn@doi [\pasp] {10.1086/379669}, \href
  {http://adsabs.harvard.edu/abs/2003PASP..115.1388H} {115, 1388}

\bibitem[\protect\citeauthoryear{{Hook}, {J{\o}rgensen}, {Allington-Smith},
  {Davies}, {Metcalfe}, {Murowinski}  \& {Crampton}}{{Hook}
  et~al.}{2004}]{hook04}
{Hook} I.~M.,  {J{\o}rgensen} I.,  {Allington-Smith} J.~R.,  {Davies} R.~L.,
  {Metcalfe} N.,  {Murowinski} R.~G.,   {Crampton} D.,  2004, \mn@doi [\pasp]
  {10.1086/383624}, \href {http://adsabs.harvard.edu/abs/2004PASP..116..425H}
  {116, 425}

\bibitem[\protect\citeauthoryear{{Ishiguro} et~al.,}{{Ishiguro}
  et~al.}{2016}]{ishiguro16}
{Ishiguro} M.,  et~al., 2016, \mn@doi [\aj] {10.3847/0004-6256/152/6/169},
  \href {http://adsabs.harvard.edu/abs/2016AJ....152..169I} {152, 169}

\bibitem[\protect\citeauthoryear{{Keller}, {Mottola}, {Skorov}  \&
  {Jorda}}{{Keller} et~al.}{2015}]{keller15}
{Keller} H.~U.,  {Mottola} S.,  {Skorov} Y.,   {Jorda} L.,  2015, \mn@doi
  [\aap] {10.1051/0004-6361/201526421}, \href
  {http://adsabs.harvard.edu/abs/2015A%26A...579L...5K} {579, L5}

\bibitem[\protect\citeauthoryear{{Knight} \& {Schleicher}}{{Knight} \&
  {Schleicher}}{2011}]{knight11b}
{Knight} M.~M.,  {Schleicher} D.~G.,  2011, \mn@doi [\aj]
  {10.1088/0004-6256/141/6/183}, \href
  {http://adsabs.harvard.edu/abs/2011AJ....141..183K} {141, 183}

\bibitem[\protect\citeauthoryear{{Knight} \& {Schleicher}}{{Knight} \&
  {Schleicher}}{2015}]{knight15a}
{Knight} M.~M.,  {Schleicher} D.~G.,  2015, \mn@doi [\aj]
  {10.1088/0004-6256/149/1/19}, \href
  {http://adsabs.harvard.edu/abs/2015AJ....149...19K} {149, 19}

\bibitem[\protect\citeauthoryear{{Knight}, {Schleicher}, {Farnham},
  {Schwieterman}  \& {Christensen}}{{Knight} et~al.}{2012}]{knight12}
{Knight} M.~M.,  {Schleicher} D.~G.,  {Farnham} T.~L.,  {Schwieterman} E.~W.,
  {Christensen} S.~R.,  2012, \mn@doi [\aj] {10.1088/0004-6256/144/5/153},
  \href {http://adsabs.harvard.edu/abs/2012AJ....144..153K} {144, 153}

\bibitem[\protect\citeauthoryear{{Kramer} \& {Noack}}{{Kramer} \&
  {Noack}}{2016}]{kramer16}
{Kramer} T.,  {Noack} M.,  2016, \mn@doi [\apjl] {10.3847/2041-8205/823/1/L11},
  \href {http://adsabs.harvard.edu/abs/2016ApJ...823L..11K} {823, L11}

\bibitem[\protect\citeauthoryear{{Landolt}}{{Landolt}}{2009}]{landolt09}
{Landolt} A.~U.,  2009, \mn@doi [\aj] {10.1088/0004-6256/137/5/4186}, \href
  {http://adsabs.harvard.edu/abs/2009AJ....137.4186L} {137, 4186}

\bibitem[\protect\citeauthoryear{{Luspay-Kuti} et~al.,}{{Luspay-Kuti}
  et~al.}{2015}]{luspaykuti15}
{Luspay-Kuti} A.,  et~al., 2015, \mn@doi [\aap] {10.1051/0004-6361/201526205},
  \href {http://adsabs.harvard.edu/abs/2015A%26A...583A...4L} {583, A4}

\bibitem[\protect\citeauthoryear{{Marcus}}{{Marcus}}{2007}]{marcus07b}
{Marcus} J.~N.,  2007, Int. Comet Quart., pp 39--66

\bibitem[\protect\citeauthoryear{{Massey} et~al.,}{{Massey}
  et~al.}{2013}]{massey13}
{Massey} P.,  et~al., 2013, in American Astronomical Society Meeting Abstracts.
  p. \#345.02

\bibitem[\protect\citeauthoryear{{Opitom}, {Snodgrass}, {Fitzsimmons}, {Jehin},
  {Manfroid}, {Tozzi}, {Faggi}  \& {Gillon}}{{Opitom} et~al.}{2017}]{opitom17}
{Opitom} C.,  {Snodgrass} C.,  {Fitzsimmons} A.,  {Jehin} E.,  {Manfroid} J.,
  {Tozzi} G.~P.,  {Faggi} S.,   {Gillon} M.,  2017, \mn@doi [\mnras]
  {10.1093/mnras/stx1591}, \href
  {http://adsabs.harvard.edu/abs/2017MNRAS.469S.222O} {469, S222}

\bibitem[\protect\citeauthoryear{{Rosenbush}, {Ivanova}, {Kiselev},
  {Kolokolova}  \& {Afanasiev}}{{Rosenbush} et~al.}{2017}]{rosenbush17}
{Rosenbush} V.~K.,  {Ivanova} O.~V.,  {Kiselev} N.~N.,  {Kolokolova} L.~O.,
  {Afanasiev} V.~L.,  2017, MNRAS, submitted

\bibitem[\protect\citeauthoryear{{Samarasinha} \& {Larson}}{{Samarasinha} \&
  {Larson}}{2014}]{samarasinha14}
{Samarasinha} N.~H.,  {Larson} S.~M.,  2014, \mn@doi [\icarus]
  {10.1016/j.icarus.2014.05.028}, \href
  {http://adsabs.harvard.edu/abs/2014Icar..239..168S} {239, 168}

\bibitem[\protect\citeauthoryear{{Schleicher}}{{Schleicher}}{2006}]{schleicher06b}
{Schleicher} D.~G.,  2006, \mn@doi [Icarus] {10.1016/j.icarus.2005.11.014},
  \href {http://adsabs.harvard.edu/abs/2006Icar..181..442S} {181, 442}

\bibitem[\protect\citeauthoryear{{Schleicher} \& {Farnham}}{{Schleicher} \&
  {Farnham}}{2004}]{schleicher04}
{Schleicher} D.~G.,  {Farnham} T.~L.,  2004, in {Festou, M.~C., Keller, H.~U.,
  \& Weaver, H.~A.} ed., , Comets II.
University of Arizona Press, Tucson, pp 449--469

\bibitem[\protect\citeauthoryear{{Schleicher}, {Millis}  \&
  {Birch}}{{Schleicher} et~al.}{1998}]{schleicher98}
{Schleicher} D.~G.,  {Millis} R.~L.,   {Birch} P.~V.,  1998, \mn@doi [Icarus]
  {10.1006/icar.1997.5902}, \href
  {http://adsabs.harvard.edu/abs/1998Icar..132..397S} {132, 397}

\bibitem[\protect\citeauthoryear{{Sekanina}}{{Sekanina}}{1991}]{sekanina91b}
{Sekanina} Z.,  1991, \jrasc, \href
  {http://adsabs.harvard.edu/abs/1991JRASC..85..324S} {85, 324}

\bibitem[\protect\citeauthoryear{{Sierks} et~al.,}{{Sierks}
  et~al.}{2015}]{sierks15}
{Sierks} H.,  et~al., 2015, \mn@doi [Science] {10.1126/science.aaa1044}, \href
  {http://adsabs.harvard.edu/abs/2015Sci...347.1044S} {347, 1044}

\bibitem[\protect\citeauthoryear{{Snodgrass}, {Tubiana}, {Bramich}, {Meech},
  {Boehnhardt}  \& {Barrera}}{{Snodgrass} et~al.}{2013}]{snodgrass13}
{Snodgrass} C.,  {Tubiana} C.,  {Bramich} D.~M.,  {Meech} K.,  {Boehnhardt} H.,
    {Barrera} L.,  2013, \mn@doi [\aap] {10.1051/0004-6361/201322020}, \href
  {http://adsabs.harvard.edu/abs/2013A\%26A...557A..33S} {557, A33}

\bibitem[\protect\citeauthoryear{{Snodgrass} et~al.,}{{Snodgrass}
  et~al.}{2016}]{snodgrass16}
{Snodgrass} C.,  et~al., 2016, \mn@doi [\aap] {10.1051/0004-6361/201527834},
  \href {http://adsabs.harvard.edu/abs/2016A%26A...588A..80S} {588, A80}

\bibitem[\protect\citeauthoryear{Snodgrass et~al.,}{Snodgrass
  et~al.}{2017}]{snodgrass17}
Snodgrass C.,  et~al., 2017, Phil.\ Trans.\ R.\ Soc.\ A, In press

\bibitem[\protect\citeauthoryear{{Taylor}, {Alexander}, {Altobelli}, {Fulle},
  {Fulchignoni}, {Gr{\"u}n}  \& {Weissman}}{{Taylor} et~al.}{2015}]{taylor15}
{Taylor} M.~G.~G.~T.,  {Alexander} C.,  {Altobelli} N.,  {Fulle} M.,
  {Fulchignoni} M.,  {Gr{\"u}n} E.,   {Weissman} P.,  2015, \mn@doi [Science]
  {10.1126/science.aaa4542}, \href
  {http://adsabs.harvard.edu/abs/2015Sci...347..387T} {347, 387}

\bibitem[\protect\citeauthoryear{{Tozzi}, {Patriarchi}, {Boehnhardt},
  {Vincent}, {Licandro}, {Kolokolova}, {Schulz}  \& {St{\"u}we}}{{Tozzi}
  et~al.}{2011}]{tozzi11}
{Tozzi} G.~P.,  {Patriarchi} P.,  {Boehnhardt} H.,  {Vincent} J.-B.,
  {Licandro} J.,  {Kolokolova} L.,  {Schulz} R.,   {St{\"u}we} J.,  2011,
  \mn@doi [\aap] {10.1051/0004-6361/201116577}, \href
  {http://adsabs.harvard.edu/abs/2011A\%26A...531A..54T} {531, A54}

\bibitem[\protect\citeauthoryear{{Vincent}, {Lara}, {Tozzi}, {Lin}  \&
  {Sierks}}{{Vincent} et~al.}{2013}]{vincent13}
{Vincent} J.-B.,  {Lara} L.~M.,  {Tozzi} G.~P.,  {Lin} Z.-Y.,   {Sierks} H.,
  2013, \mn@doi [\aap] {10.1051/0004-6361/201219350}, \href
  {http://adsabs.harvard.edu/abs/2013A\%26A...549A.121V} {549, A121}

\bibitem[\protect\citeauthoryear{{Vincent} et~al.,}{{Vincent}
  et~al.}{2015}]{vincent15}
{Vincent} J.-B.,  et~al., 2015, \mn@doi [\nat] {10.1038/nature14564}, \href
  {http://adsabs.harvard.edu/abs/2015Natur.523...63V} {523, 63}

\bibitem[\protect\citeauthoryear{{Vincent} et~al.,}{{Vincent}
  et~al.}{2016a}]{vincent16b}
{Vincent} J.-B.,  et~al., 2016a, \mn@doi [\mnras] {10.1093/mnras/stw2409},
  \href {http://adsabs.harvard.edu/abs/2016MNRAS.462S.184V} {462, S184}

\bibitem[\protect\citeauthoryear{{Vincent} et~al.,}{{Vincent}
  et~al.}{2016b}]{vincent16a}
{Vincent} J.-B.,  et~al., 2016b, \mn@doi [\aap] {10.1051/0004-6361/201527159},
  \href {http://adsabs.harvard.edu/abs/2016A%26A...587A..14V} {587, A14}

\bibitem[\protect\citeauthoryear{{Weiler}, {Rauer}  \& {Helbert}}{{Weiler}
  et~al.}{2004}]{weiler04}
{Weiler} M.,  {Rauer} H.,   {Helbert} J.,  2004, \mn@doi [\aap]
  {10.1051/0004-6361:20031610}, \href
  {http://adsabs.harvard.edu/abs/2004A%26A...414..749W} {414, 749}

\bibitem[\protect\citeauthoryear{{Zacharias}, {Finch}, {Girard}, {Henden},
  {Bartlett}, {Monet}  \& {Zacharias}}{{Zacharias} et~al.}{2013}]{zacharias13}
{Zacharias} N.,  {Finch} C.~T.,  {Girard} T.~M.,  {Henden} A.,  {Bartlett}
  J.~L.,  {Monet} D.~G.,   {Zacharias} M.~I.,  2013, \mn@doi [\aj]
  {10.1088/0004-6256/145/2/44}, \href
  {http://adsabs.harvard.edu/abs/2013AJ....145...44Z} {145, 44}

\bibitem[\protect\citeauthoryear{{Zaprudin}, {Lehto}, {Nilsson}, {Somero},
  {Pursimo}, {Snodgrass}  \& {Schulz}}{{Zaprudin} et~al.}{2017}]{zaprudin17}
{Zaprudin} B.,  {Lehto} H.~J.,  {Nilsson} K.,  {Somero} A.,  {Pursimo} T.,
  {Snodgrass} C.,   {Schulz} R.,  2017, A\&A, submitted

\makeatother
\end{thebibliography}



\bsp	
\label{lastpage}
\end{document}